\pdfoutput=1
\documentclass[11pt,a4paper]{article}

\usepackage[T1]{fontenc}
\usepackage[margin=2.5cm]{geometry}
\usepackage{amsmath, amsfonts, amssymb, amsthm, mathtools}
\usepackage{graphicx}
\usepackage{microtype}
\usepackage{xcolor}
\usepackage{todonotes}
\usepackage{authblk}
\usepackage[hidelinks]{hyperref}
\usepackage{orcidlink}
\usepackage[capitalise]{cleveref}

\newif\ifjournal
\newcommand{\journalbreakyes}{\\&}
\newcommand{\journalbreak}{\ifjournal\expandafter\journalbreakyes\fi}

\newcommand{\np}{\textsf{NP}}
\newcommand{\p}{\textsf{P}}
\newcommand{\fpt}{\textsf{FPT}}
\newcommand{\w}[1]{\textsf{W[#1]}}
\newcommand{\wone}{\w{1}}

\newcommand{\ethy}{\textsf{ETH}}
\newcommand{\integers}{\ensuremath{\mathbb{N}_{\ge 0}}}
\newcommand{\naturals}{\ensuremath{\mathbb{N}_{>0}}}
\newcommand{\reals}{\ensuremath{\mathbb{R}}}
\newcommand{\bigO}{\ensuremath{\mathcal{O}}}

\newcommand{\switch}{\ensuremath{s}}
\newcommand{\sgraph}{\ensuremath{\widehat{G}}}
\newcommand{\switches}{\ensuremath{\mathcal{S}}}
\newcommand{\x}[2]{\ensuremath{#1 \circ #2}}

\newcommand{\config}{\ensuremath{c}}
\newcommand{\configs}{\ensuremath{\mathcal{C}}}
\newcommand{\family}{\ensuremath{\Pi}}

\newcommand{\prob}[5]{
\begin{samepage}
    \begin{description}
\setlength\topsep{-.15ex} \setlength\itemsep{-.2ex}
        \item[#1]
        \item[#2]#3
        \item[#4]#5
    \end{description}
\end{samepage}
}
\newcommand{\probname}[1]{{\normalfont\textsc{#1}}}
\newcommand{\probdef}[3]{\prob{\probname{#1}}{Input:}{#2}{Question:}{#3}}

\newcommand{\tcp}{\probname{Tran\-si\-tion-Com\-pat\-i\-ble Path}}
\newcommand{\psgi}{\probname{Partitioned Subgraph Isomorphism}}
\newcommand{\lcs}{\probname{Local-Con Switching}}
\newcommand{\wlcs}{\probname{Weighted Local-Con Switching}}
\newcommand{\oolcs}{\probname{On-Off Local-Con Switching}}
\newcommand{\gcs}{\probname{Global-Con Switching}}
\newcommand{\wgcs}{\probname{Weighted Global-Con Switching}}
\newcommand{\oogcs}{\probname{On-Off Global-Con Switching}}
\newcommand{\sat}{\probname{SAT}}
\newcommand{\wsat}{\probname{WSAT}}

\theoremstyle{plain}
\newtheorem{theorem}{Theorem}
\newtheorem{lemma}[theorem]{Lemma}
\newtheorem{corollary}[theorem]{Corollary}
\theoremstyle{definition}
\newtheorem{definition}[theorem]{Definition}
\newtheorem{construction}[theorem]{Construction}
\ifdefined\AddToHook
  \AddToHook{env/lemma/begin}{\crefalias{theorem}{lemma}}
  \AddToHook{env/corollary/begin}{\crefalias{theorem}{corollary}}
  \AddToHook{env/definition/begin}{\crefalias{theorem}{definition}}
  \AddToHook{env/construction/begin}{\crefalias{theorem}{construction}}
\fi
\crefname{construction}{Construction}{Constructions}

\definecolor{chartreuse3}{rgb}{0.4, 0.804, 0}
\definecolor{dodgerblue3}{rgb}{0.094, 0.455, 0.804}
\definecolor{deeppink3}{rgb}{0.804, 0.063, 0.463}

\title{Algorithms and Complexity of Connectivity Switching in Graphs\thanks{This is the full version of a paper presented at LATIN 2026. This work is based on Christoph Kern's master thesis~\cite{Kern25}. Alexander Firbas acknowledges support from the Vienna Science and Technology Fund (WWTF) [10.47379/ICT22029] and the Austrian Science Fund (FWF) [10.55776/Y1329].}}

\author{Christoph Kern~\orcidlink{0000-0001-6308-502X}}
\author{Alexander Firbas~\orcidlink{0009-0007-2049-2144}}
\author{Manuel Sorge~\orcidlink{0000-0001-7394-3147}}
\affil{TU Wien, Austria\\
  \texttt{christoph@kern.codes}, \texttt{\{afirbas,manuel.sorge\}@ac.tuwien.ac.at}}
\date{}

\begin{document}

\maketitle

\begin{abstract}
We study computational problems in switch graphs.
A \emph{switch graph} consists of a vertex set and a set of \emph{switches}, which are collections of edge sets.
The goal is to find a \emph{configuration}---selecting one edge set per switch---such that the resulting graph satisfies a desired property.
This general framework unifies several switching models from prior~work, with applications in solving Boolean equation systems, network routing, and SNP genotyping.
We focus on connectivity-related switching problems, such as finding a configuration that yields a connected graph (global connectivity) or that connects a specific pair of vertices (local connectivity), presenting polynomial-time algorithms and establishing \np{}-completeness results depending on the switch structure.
For local connectivity, we obtain a full complexity dichotomy.
Furthermore, we introduce a natural weighted variant and analyze its parameterized complexity, proving \wone-hardness with respect to the solution weight via reductions from restricted variants of the \emph{weighted satisfiability problem} (\probname{WSAT}).
As a byproduct, we resolve an open question of Kanj and Szeider~\ifjournal(TCS 2015)\else\cite{Kanj15}\fi{} on the parameterized complexity of a specific \probname{WSAT} variant.

\end{abstract}

\section{Introduction}
\label{sec:intro}
In various applications we are given an interconnection network whose structure we can control by making small changes and we aim to achieve certain goals.
For instance, in brewery valve systems we want to connect certain terminal pairs by opening or closing valves~\cite{Huckenbeck93,Huckenbeck97}, in train-track systems we want to enable trains to reach certain points by configuring railroad switches~\cite{Reinhardt09}, in graph clustering we want to add or delete noisy edges to obtain a clean clustering~\cite{Bocker13}, and so on~\cite{Meinel89,Szeider03,Groote09,Katz12}.
Abstractly, we can model these problems as follows.
We are given a so-called switch graph and we want to find a configuration of the switches such that the resulting graph satisfies a desired property.

\begin{figure}[t]
    \centering
    \includegraphics[page=15]{graphics/figures.pdf}
    \caption{A switch graph $\sgraph$ with $V(\sgraph) = \{v_0,v_1,v_2,v_3,v_4\}$ and $\switches(\sgraph)=\{\textcolor{dodgerblue3}{\switch_0}, \textcolor{chartreuse3}{\switch_1}, \textcolor{deeppink3}{\switch_2}\}$, and a configuration $\config$ of $\sgraph$ such that $\config(\textcolor{dodgerblue3}{\switch_0})=1$, $\config(\textcolor{chartreuse3}{\switch_1})=1$, and $\config(\textcolor{deeppink3}{\switch_2})=2$.}
    \label{fig:example-switch-graph}
\end{figure}

Formally, a switch graph \sgraph{} consists of a set~$V(\sgraph)$ of vertices and a set~$\switches(\sgraph)$ of \emph{switches}.
Each switch $\switch \in \switches(\sgraph)$ has $|\switch| \in \naturals$ \emph{positions}, indexed by integers $0 \le i < |\switch|$, where each position $\switch[i]$ is a set of edges $\switch[i] \subseteq \binom{V(\sgraph)}{2}$.
(Note that ordinary edges can be modeled by trivial switches.)
A \emph{configuration} of \sgraph{} is a function~$\config \colon \switches(\sgraph) \to \integers$ that maps each switch~$\switch \in \switches(\sgraph)$ to exactly one switch position~$i < |\switch|$. See \cref{fig:example-switch-graph} for an example switch graph and configuration.

Inspired by the applications, we focus here on finding a configuration that ensures local connectivity (\lcs), that is, to connect two given vertices $a, b$, or global connectivity (\gcs), that is, to pairwise connect all vertices.
In general, both problems are \np-hard, as we show, but there are some efficiently solvable cases for restricted types of switches~\cite{Katz12}.

The goal in this work is two-fold.
First, we want to determine how the form of the switches influences the complexity of \lcs\ and \gcs. 
That is, for which types of switches are they still polynomial-time solvable?
Here, we mainly distinguish two features: whether the switches are connected (their union of positions induces a connected graph), and the number of edges in each position.
For instance, we refer to switches with two positions, one with at most two edges and one with at most one edge as \emph{$\{2, 1\}$-bounded}.
Similarly, a \emph{$\{1^\star\}$-bounded} switch has an arbitrary number of positions, each with at most one edge. (See \cref{sec:prelims} for details.)
Second, we want to study the influence of weights on the complexity of the two problems:
It is natural to consider settings in which the switches have a specified on- and off-position and we can afford to switch only at most a given number $\lambda$ many switches to the on-position.
More generally, we can equip switch positions with weights and look for configurations that have at most a given weight.
We refer to these two settings by the \textsc{On-Off} and \textsc{Weighted} qualifiers, respectively.\ifjournal\else\looseness=-1\fi

We achieve a series of dichotomy results as follows (see also \cref{fig:results-overview}):
Let us focus on \lcs\ first.
Note that the problem becomes trivial when restricting to $\{1, 0\}$-bounded switches.
However, even the first nontrivial case, namely $\{1, 1\}$-bounded switches, makes \lcs\ \np-complete.
In fact, we obtain a sharp dichotomy in the $\{1^\star\}$-bounded case: \lcs\ is in \p{} if each switch induces a complete multipartite graph and otherwise \np\nobreakdash-complete~(\cref{thm:dichotomy-lcs}).
When restricting to connected switches, Katz et al.~\cite{Katz12} showed that $\{1, 1\}$-bounded switches make \lcs\ polynomial-time solvable.
We match this with a lower bound where $\{2, 1\}$-bounded connected switches yield \np-completeness~(\cref{thm:hardness-lcs-star-switches}).

The \textsc{On-Off} and \textsc{Weighted} cases of \lcs\ are more general and hence we broaden our notion of tractability:
\oolcs\ can be solved in $|I|^{\lambda + \bigO(1)}$ time by simply trying all possibilities of turning at most $\lambda$ switches on, where $|I|$ is the instance size.
It would be interesting to improve this to a fixed-parameter running time $f(\lambda) \cdot |I|^{\bigO(1)}$.
Alas, we show that even in the first nontrivial case of $\{1, 1\}$-bounded switches and for $\{2, 1\}$-bounded connected switches, \oolcs\ is unlikely to be fixed-parameter tractable~(\cref{thm:wlcs-11-switches,thm:wlcs-star-switches}).
To prove this we show a relation between series-parallel switch graphs and Boolean formulas.
Partly we can then rely on known hardness results for weighted satisfiability problems to prove hardness for \oolcs.
However, for $\{1, 1\}$-bounded switches, no corresponding hard satisfiability variant was known.
Here we provide a new hardness reduction (\cref{thm:psi-to-wsat-11}) and thereby answer an open question posed by Kanj and Szeider~\cite{Kanj15}.
This result may be of independent interest.

\begin{figure}[t]
    \centering
    \includegraphics[page=19,scale=0.85]{graphics/figures.pdf}
    \caption{A Hasse diagram of major structural constraints on the switches, with an arrow from $A$ to $B$ indicating that $A$ is a special case of $B$. Complexities are given for each constraint in the unweighted, weighted, and on-off cases for both \lcs\ and \gcs. Note that both \wone- and \w{SAT}-hardness indicate that the problem is unlikely to be fixed-parameter tractable.}
    \label{fig:results-overview}
\end{figure}

For \gcs, Katz et al.~\cite{Katz12} observed that there is a strong relation to matroids and, using this, showed that \gcs\ is polynomial-time solvable for $\{1^\star\}$-bounded star-shaped switches.
We generalize their approach to arbitrary $\{1^\star\}$-bounded switches, also in the weighted case~(\cref{thm:wgcs-matroid}), and, in the unweighted case, to all switches whose positions form a matroid~(\cref{thm:gcs-matroid-switches}).
To show the limits of this idea, we also provide a matching lower bound: The simplest switches that are not necessarily matroid are $\{2, 1\}$-bounded, and already with such switches \gcs\ (\cref{thm:gcs-21-npc}) and the \textsc{On-Off} variant (\cref{thm:wgcs-star-switches}) become hard, even when restricted to connected switches.

\section{Preliminaries}
\label{sec:prelims}

A simple graph $G$ consists of a set $V(G)$ of vertices and a set $E(G) \subseteq \binom{V(G)}{2}$ of edges.
For better readability, we often represent an edge $\{u,v\} \in E(G)$ as~$uv$.
For a vertex $v \in V(G)$, let $N_G(v) \coloneqq \{u \in V(G) \mid uv \in E(G)\}$ denote its neighbors, and $\deg_G(v) \coloneqq |N_G(v)|$ the \emph{vertex degree}.
All paths in this work are simple, that is, without repeated vertices or edges.
For other graph-theoretic notions not defined here, we refer to Diestel~\cite{Diestel12}.

\subsection{Switch Graphs}

Recall from \cref{sec:intro} that a switch graph~\sgraph{} consists of a vertex set~$V(\sgraph)$ and a set~$\switches(\sgraph)$ of switches, where each switch~$\switch \in \switches(\sgraph)$ has $|\switch|$ positions $\switch[i] \subseteq \binom{V(\sgraph)}{2}$ for $i < |\switch|$. A configuration of \sgraph{} is a function~$\config$ that maps each switch~$\switch$ to one of its positions $\config(\switch) < |\switch|$, and we denote the set of all configurations of \sgraph{} by $\configs(\sgraph)$. Each configuration~$\config$ realizes a simple graph $\x{\sgraph}{\config}$ (read ``\sgraph{} configured by \config'') with vertex set $V(\x{\sgraph}{\config}) \coloneqq V(\sgraph)$ and edge set $E(\x{\sgraph}{\config}) \coloneqq \bigcup_{\switch \in \switches(\sgraph)} \switch[\config(\switch)]$, that is, the union of the edge sets of the selected switch positions.

To better classify the structure of a switch~$\switch$, we commonly look at its graph structure, provided by the \emph{switch-induced graph} $G[\switch]$, with edge set $E(G[\switch]) \coloneqq \bigcup_{i < |\switch|} \switch[i]$ containing all edges that appear in at least one position of $\switch$, and vertex set $V(G[\switch]) \coloneqq \bigcup_{e \in E(G[\switch])} e$ consisting of all vertices that are incident to at least one edge in $E(G[\switch])$. Given a family of graphs $\family$, we say that a switch~$\switch$ is \emph{$\family$-shaped} if $G[\switch]$ is isomorphic to a graph $G \in \family$. Common graph families we consider are complete multipartite graphs $\mathcal{K}_\mathrm{multi}$, and star graphs $\mathcal{K}_{1,\star}$.

We also classify switches based on the sizes of their positions. For a multiset~$\Sigma$ of positive integers, we say that a switch $\switch$ is \emph{$\Sigma$-bounded} if there exists an injective mapping $\pi\colon \{0, 1, \dots, |\switch|-1\} \to \Sigma$ such that for all $i < |\switch|$, we have $|\switch[i]| \le \pi(i)$. We use exponent notation to denote multiplicities in $\Sigma$ and $\star$ to denote an arbitrarily large number. For example, a $\{\star^2\}$-bounded switch $\switch$, also called an \emph{on-off switch}, has at most two positions---$\switch[0]$ is called the \emph{off-position}, $\switch[1]$ is called the \emph{on-position}---each of arbitrary size, while a $\{1^\star\}$-bounded switch has arbitrarily many positions, each containing at most one edge. Furthermore, a $\{1\}$-bounded $\{K_2\}$-shaped switch is called a \emph{static-edge switch}, as it acts like a permanent edge appearing in every configuration of a switch graph.

A switch graph \sgraph{} can also be seen as a multigraph with edge set $E(\sgraph) \coloneqq \biguplus_{\switch \in \switches(\sgraph)} E(G[\switch])$, where each edge $e \in E(\sgraph)$ is associated with exactly one switch~$\switch \in \switches(\sgraph)$ and a nonempty set $P$ of its switch positions. Let~$n \coloneqq |V(\sgraph)|$, $m \coloneqq |E(\sgraph)|$, and $\sigma \coloneqq |\switches(\sgraph)|$.

Note that all problem variants considered in this paper ask for a configuration of a given switch graph satisfying a property that can be verified in polynomial time. Consequently, all \np-hard problems presented in this paper are also \np-complete.

\subsection{Parameterized Complexity}

We refer to standard textbooks~\cite{Cygan15,Flum06,Downey13} for an introduction to parameterized complexity theory, and assume the reader is familiar with basic concepts such as fixed-parameter tractability, parameterized reductions, and the \textsf{W}-hierarchy, in particular \wone{} and \w{SAT}.

We call a parameterized reduction~\cite[Definition~13.1]{Cygan15} from $A$ to $B$ \emph{parameter-linear} if it maps an instance $(x,k)$ of $A$ to an instance $(x', k')$ of $B$ with parameter $k' \in \bigO(k)$. Clearly, parameter-linear reductions are closed under composition. By~\cite[Observation~14.22]{Cygan15}, if $B$ can be solved in time $f'(k') \cdot |x'|^{o(h(k'))}$, then $A$ can be solved in time $f(k) \cdot |x|^{o(h(k))}$, where $f$ and $f'$ are computable functions and $h$ is a concave function such as $k$, $k/\log k$, or $\sqrt{k}$.

The following well-known problem serves as the source of our \wone-hardness reductions:

\begin{definition}
    \label{def:psi}
    \probdef{\psgi{}}{
    Two simple graphs $G$ and $H$, a partition $\mathcal{P} \coloneqq \{P_x \mid x \in V(H)\}$ of~$V(G)$ with $P_x \neq \emptyset$ for all $x \in V(H)$.
}{
    Is there a mapping $\phi\colon V(H) \to V(G)$ such that $\phi(x) \in P_x$ for all~$x \in V(H)$ and $\phi(x)\phi(y) \in E(G)$ for all $xy \in E(H)$?
}
\end{definition}

In other words, the problem asks whether $H$ is isomorphic to a subgraph of $G$ such that each vertex $x \in V(H)$ is mapped to a vertex in the corresponding part $P_x$ of the partition~$\mathcal{P}$. Observe that if $H$ is a complete graph, this is exactly \probname{Multicolored Clique}, which is known to be \wone-hard parameterized by $|V(H)|$ \cite[Theorem 13.25]{Cygan15}. Since $|E(H)| \le |V(H)|^2$, we obtain that \psgi{} is \wone-hard parameterized by $k \coloneqq |E(H)|$.

Marx showed that, assuming the \emph{Exponential Time Hypothesis} (\ethy{}), \psgi{} has no $f(k) \cdot n^{o\left(k / \log k\right)}$-time algorithm, where $n \coloneqq |V(G)|$ \cite[Corollary~6.3]{Marx10}. Since \ethy{} has recently been claimed to be false \cite{OpenAI26,Anthropic26}, we state our results as parameter-linear reductions from \psgi{}, so that any such lower bound for \psgi{} carries over. Without \ethy{}, the best bound we are aware of is $f(k) \cdot n^{o\left(\sqrt{k}\right)}$, which holds under the $k \times k$ Clique Hypothesis recently proposed by Szeider~\cite{Szeider26}.

\section{Local Connectivity}
\label{sec:stcon}

In this section, we investigate configurations that connect two distinguished vertices $a$ and $b$ by a path, formally defined in the following decision problem:

\begin{definition}
    \probdef{\lcs{}}{
        A switch graph \sgraph{} and two distinct vertices $a,b \in V(\sgraph{})$.
    }{
        Is there a configuration $\config \in \configs(\sgraph)$ such that $a$ and $b$ are connected in~$\x{\sgraph}{\config}$?
    }
\end{definition}

\subsection{Connection to Transition-Compatible Paths}
\begin{figure}[t]
    \centering
    \includegraphics[page=11]{graphics/figures.pdf}
    \caption{Example of a graph $G$ with transition system $T$ (left) and the corresponding switch graph $\sgraph_T$ (right), constructed according to \cref{con:transition-graph-to-switch}.}
    \label{fig:transition-graph-to-switch}
\end{figure}
Szeider~\cite{Szeider03} introduced a problem called \tcp{}, which is closely related to the \lcs{} problem studied in this paper.
Given a graph~$G$, let $T[v]$ be a transition graph for a vertex $v \in V(G)$, whose vertices are the edges of $G$ incident to $v$. Two such edges are adjacent in $T[v]$ if they form an \emph{allowed transition} via $v$. A \emph{transition system} $T$ is the collection of all transition graphs $T[v]$ for every vertex $v \in V(G)$. A path $P \coloneqq \langle v_0, v_1, \dots, v_k \rangle$ in $G$ is called \emph{$T$-compatible} if it consists only of allowed transitions; that is, for each $i \in \{1,2,\dots,k-1\}$, edge~$v_{i-1}v_i$ is adjacent to edge~$v_iv_{i+1}$ in $T[v_i]$. This leads to the following problem definition:

\begin{definition}
    \probdef{\tcp{}}{
    A graph $G$ with transition system $T$, two distinct vertices $a,b \in V(G)$.
}{
    Is there a $T$-compatible path from $a$ to $b$ in $G$?
}
\end{definition}

We can reduce \tcp{} to \lcs{} restricted to $\{1^\star\}$-bounded switches, and vice versa. In the left-to-right direction, we use the following construction, illustrated in \cref{fig:transition-graph-to-switch} by an example:

\begin{construction}
    \label{con:transition-graph-to-switch}
    Let $G$ be a graph with transition system $T$, and $a,b \in V(G)$ be two distinct vertices. We construct a corresponding switch graph $\sgraph_T$ as follows:
    \begin{itemize}
        \item Let $V(\sgraph_T) \coloneqq \{a,b\} \cup E(G)$.
        \item For each $v \in V(G) \setminus \{a,b\}$, we create a $\{1^\star\}$-bounded switch $\switch_v$ such that $E(G[\switch_v]) = E(T[v])$.
        \item For each edge $e$ incident to a terminal $v \in \{a,b\}$, we introduce a static-edge switch $\switch_e$ with $\switch_e[0] \coloneqq \{ve\}$.
    \end{itemize}
\end{construction}

Intuitively, a $T$-compatible path from $a$ to $b$ in $G$ can be seen as a sequence of allowed transitions between pairs of adjacent edges, with at most one transition per vertex.
(Recall that all our paths are simple.)
In $\sgraph_T$, each vertex~$v \in V(G)$ and its transition graph $T[v]$ are represented by a $\{1^\star\}$-bounded switch $\switch_v$, where each position corresponds to one such allowed transition. Configuring the switches according to these transitions yields a path from $a$ to $b$ in~$\sgraph_T$ following the same sequence. The converse can be argued similarly. We thus obtain:

\begin{lemma}
    \label{lem:transition-graph-to-switch}
    Let $G$ be a graph with transition system $T$, and $a,b \in V(G)$ be two distinct vertices. Then, there exists a $T$-compatible path from $a$ to $b$ in $G$ if and only if $\sgraph_T$, as constructed by \cref{con:transition-graph-to-switch}, admits a path from $a$ to $b$ under some configuration $\config \in \configs(\sgraph_T)$.
\end{lemma}
\begin{proof}
We show both directions of the equivalence separately.
\begin{description}
    \item[$(\Rightarrow)$] Let $P \coloneqq \langle v_0, v_1, \dots, v_{k-1}, v_k \rangle$ be a $T$-compatible path from $a=v_0$ to $b=v_k$ in $G$. For each inner vertex $v_i$ with $i \in \{1,2,\dots,k-1\}$, the pair $\{v_{i-1}v_i, v_iv_{i+1}\}$ forms an allowed transition in $T[v_i]$. We define a configuration $\config$ of $\sgraph_T$ by setting each corresponding switch $\switch_{v_i}$ to the position that includes edge $\{v_{i-1}v_i, v_iv_{i+1}\}$. All other switches are set to position $0$. Then, the sequence $P' \coloneqq \langle v_0, v_{0}v_1, v_1v_2, \dots, v_{k-1}v_k, v_k \rangle$ forms a path from $a$ to $b$ in $\sgraph_T$ under configuration $\config$, since $P$ is a path and therefore no edge transition is repeated, which implies that no edge vertex is revisited in $P'$.

    \item[$(\Leftarrow)$] Let $\config$ be a configuration of $\sgraph_T$ that admits a path $P' \coloneqq \langle a, e_1, e_2, \dots, e_k,\allowbreak b \rangle$. For each pair of consecutive edge vertices $\{e_i, e_{i+1}\}$ in $P'$, we know by construction that this pair shares a common endpoint $v \in V(G)$, and that it forms a valid transition in $T[v]$ since $\{e_i, e_{i+1}\} \in \switch_v[\config(\switch_v)]$. This implies that a path may transition from edge $e_i$ to $e_{i+1}$ via $v$ in a transition-compatible path in $G$. As $\{e_i, e_{i+1}\}$ is the only edge contributed by switch $\switch_v$ in configuration $\config$, it is also the only transition via $v$ of any pair of consecutive edge vertices in $P'$. Therefore, the sequence $\langle e_1, e_2, \dots, e_k \rangle$ of edges forms a $T$-compatible path, as this prevents a single vertex from being visited multiple times. Finally, since $e_1$ is incident to $a$ and $e_k$ to $b$ by the definition of the static edges $\{a, e_1\}$ and $\{e_k, b\}$, we conclude that $G$ admits a $T$-compatible path from $a$ to $b$.\qedhere
\end{description}
\end{proof}

For reducing from \lcs\ to \tcp\ we use the following construction, illustrated in \cref{fig:switch-to-transition-graph} by an example:

\begin{figure}[t]
    \centering
    \includegraphics[page=13]{graphics/figures.pdf}
    \caption{Example of a switch graph $\sgraph$ (left) and the corresponding graph $G$ with transition system $T$ (right), constructed according to \cref{con:switch-to-transition-graph}.}
    \label{fig:switch-to-transition-graph}
\end{figure}

\begin{construction}
    \label{con:switch-to-transition-graph}
    Let $\sgraph$ be a switch graph restricted to $\{1^\star\}$-bounded switches and $a,b \in V(\sgraph)$ be two distinct vertices. We construct a corresponding graph $G$ with transition system $T$ as follows:
    \begin{itemize}
        \item Let $V(G)$ consist of the original vertices $V(\sgraph)$ and a switch vertex $v_\switch$ for each $\switch \in \switches(\sgraph)$.
        \item Let $E(G) \coloneqq \bigcup_{\switch \in \switches(\sgraph)} \{v_\switch u \mid u \in V(G[\switch])\}$.
        \item For each switch vertex $v_\switch \in V(G)$, let $T[v_\switch]$ be a transition graph such that $E(T[v_\switch]) \coloneqq \{\{uv_\switch, wv_\switch\} \mid uw \in E(G[\switch])\}$.
        \item For each original vertex $v \in V(\sgraph)$, let $T[v]$ be a complete graph, that is, $E(T[v]) \coloneqq \binom{V(T[v])}{2}$.
    \end{itemize}
\end{construction}

Observe that every edge $uw$ in a switch $\switch$ corresponds to an allowed transition in $T[v_\switch]$, enabling a connection between $u$ and $w$ through the switch vertex $v_\switch$. Intuitively, this allows any path in a realized graph of \sgraph{} to be transformed into a $T$-compatible path in $G$. The converse direction follows by an analogous reasoning. This leads to the following lemma:

\begin{lemma}
    \label{lem:switch-to-transition-graph}
    Let $\sgraph$ be a switch graph restricted to $\{1^\star\}$-bounded switches. Then, there exists a path from $a$ to $b$ in $\sgraph$ under some configuration $\config \in \configs(\sgraph)$ if and only if the graph $G$ constructed by \cref{con:switch-to-transition-graph} admits a $T$-compatible path from $a$ to $b$.
\end{lemma}
\begin{proof}
We show both directions of the equivalence separately.
\begin{description}
    \item[$(\Rightarrow)$] Let $\config$ be a configuration of \sgraph{} that admits a path $P \coloneqq \langle v_0,v_1,\dots,v_k \rangle$ from $a=v_0$ to $b=v_k$. Since every switch adds at most one edge to $\x \sgraph \config$, each consecutive pair of vertices $v_i, v_{i+1}$ in $P$ can be mapped to a distinct switch $\switch_i \in \switches(\sgraph)$ where $v_iv_{i+1} \in E(G[\switch_i])$. By construction, the pair $\{v_iv_{\switch_i}, v_{i+1}v_{\switch_i}\}$ is an allowed transition in $T[v_{\switch_i}]$. Thus, $P' \coloneqq \langle v_0, v_{\switch_0}, v_1, v_{\switch_1}, \dots, v_{k-1}, v_{\switch_{k-1}}, v_k \rangle$ is a $T$-compatible path from $a$ to $b$ in $G$.

    \item[$(\Leftarrow)$] Let $P'$ be a $T$-compatible path from $a=v_0$ to $b=v_k$ in $G$. Since every switch vertex is only adjacent to original vertices and vice versa, the path $P' \coloneqq \langle v_0, v_{\switch_0}, v_1, v_{\switch_1}, \dots, v_{k-1}, v_{\switch_{k-1}}, v_k \rangle$ alternates between original vertices and switch vertices. Therefore, for each switch vertex $v_{\switch_i}$ within the path, the pair $\{v_iv_{\switch_i}, v_{i+1}v_{\switch_i}\}$ must be an allowed transition in $T[v_{\switch_i}]$. By the construction of $G$, this implies that $v_iv_{i+1} \in E(G[\switch_i])$. We create a configuration $\config$ of \sgraph{} such that $v_iv_{i+1} \in \switch_i[\config(\switch_i)]$ for each $i < k$. Then, path $P \coloneqq \langle v_0, v_1, \dots, v_k \rangle$ is a path from $a$ to $b$ in $\x \sgraph \config$.\qedhere
\end{description}
\end{proof}

Beyond the equivalence with \tcp{} established above, we further require a transformation for the upcoming complexity results: replacing $\{1^\star\}$-bounded switches of arbitrary shape with $\mathcal{K}_{1,\star}$-shaped ones without changing which pairs of vertices can be connected. The following construction achieves this and is illustrated in \cref{fig:1-switch-to-star-switches} by two examples:

\begin{figure}[t]
    \centering
    \includegraphics[page=12]{graphics/figures.pdf}
    \caption{Transforming $\{1^\star\}$-bounded switches (left: $K_3$-shaped, right: $L_{3,1}$-shaped) into $\mathcal{K}_{1,\star}$-shaped switches, allowing for the same connections between original vertices while preventing multiple connections to occur simultaneously in a path of a realized graph.}
    \label{fig:1-switch-to-star-switches}
\end{figure}

\begin{construction}
    \label{con:1-switch-to-star-switches}
    Let \sgraph{} be a switch graph and $\switch \in \switches(\sgraph)$ a $\{1^\star\}$-bounded switch. We construct an equivalent switch graph $\sgraph'$ by introducing a fresh vertex $w$ and replacing $\switch$ by a set of new switches whose edges are all incident to $w$.
    
    Let $\{V_0, \dots, V_{k-1}\}$ be a partition of $V(G[\switch])$ such that for every pair of indices $i \ne j$, all possible edges between vertices in $V_i$ and vertices in $V_j$ are contained in $E(G[\switch])$. For each index $i < k$, we introduce a new switch $\switch_i$ to locally model the structure of $G[\switch][V_i]$.

    Let $\{V_{i,0}, \dots, V_{i,{\ell-1}}\}$ be an \emph{edge clique cover} of $G[\switch][V_i]$, that is, a collection of subsets of $V_i$ such that, for each $j < \ell$, the induced subgraph $G[\switch][V_{i,j}]$ is a clique, and every vertex and edge in $G[\switch][V_i]$ is contained in at least one of the $V_{i,j}$. For each subset $V_{i,j}$, we define a switch position $\switch_i[j] = \{wv \mid v \in V_{i,j}\}$.
\end{construction}

Intuitively, the construction ensures that for any pair of vertices $u,v \in V(G[\switch])$, the original switch $\switch$ can be configured to connect $u$ and $v$ if and only if the new switches can be configured to connect them via the fresh vertex $w$. Since $w$ can be traversed at most once in any path, only one such connection can be used, thereby replicating the behavior of the original $\{1^\star\}$-bounded switch $\switch$. The correctness of \cref{con:1-switch-to-star-switches} is established by the following lemma:

\begin{lemma}
    \label{lem:1-switch-to-star-switches}
    Let \sgraph{} be a switch graph and let $\switch \in \switches(\sgraph)$ be a $\{1^\star\}$-bounded switch. Then, \sgraph{} admits a path from $a$ to $b$ under some configuration if and only if the transformed switch graph $\sgraph'$, obtained via \cref{con:1-switch-to-star-switches}, also admits a path from $a$ to $b$ under some configuration.
\end{lemma}
\begin{proof}
    Let $\{V_0, \dots, V_{k-1}\}$ be a partition of $V(G[\switch])$ used in the construction of $\sgraph'$. We prove both directions of the equivalence separately.

    \begin{description}
        \item[$(\Rightarrow)$] Let $\config$ be a configuration of \sgraph{} that admits a path $P$ from $a$ to $b$. We extend it to a configuration $\config'$ of $\sgraph'$ as follows. If $P$ does \emph{not} use any edge from $\switch$, then $\config'$ can assign arbitrary positions to the new switches in $\sgraph'$, and $P$ remains a valid path from $a$ to $b$ in $\x {\sgraph'} {\config'}$.

        Otherwise, let $uv \in \switch[\config(\switch)]$ be the edge used by $P$. If $u \in V_i$ and $v \in V_j$ for some $i \ne j$, then we configure $\switch_i$ such that $uw \in \switch_i[\config'(\switch_i)]$ and $\switch_j$ such that $wv \in \switch_j[\config'(\switch_j)]$. The remaining new switches can be configured arbitrarily. Replacing $uv$ in $P$ with the subpath $u \to w \to v$ yields a valid path from $a$ to $b$ in $\x {\sgraph'} {\config'}$.

        If instead $u,v \in V_i$ for some $i < k$, then by the edge clique cover $\{V_{i,0}, \dots, V_{i,\ell-1}\}$ used in the construction, there exists a $V_{i,j}$ such that $\{u,v\} \subseteq V_{i,j}$. We set $\config'(\switch_i) = j$ so that $uw, wv \in \switch_i[j]$. Again, replacing $uv$ in $P$ by $u \to w \to v$ yields a path from $a$ to $b$ in $\x {\sgraph'} {\config'}$.

        \item[$(\Leftarrow)$] Let $\config'$ be a configuration of $\sgraph{}'$ that admits a path $P$ from $a$ to $b$. We extend $\config'$ to a configuration $\config$ of \sgraph{} as follows. If $P$ does not traverse the fresh vertex $w$, then it avoids all edges added in place of $\switch$, and $\switch$ can be configured arbitrarily. Hence, $P$ is also a path from $a$ to $b$ in $\x \sgraph \config$.

        Otherwise, let $u$ and $v$ be the neighbors of $w$ in $P$. If $uw \in \switch_i[\config'(\switch_i)]$ and $wv \in \switch_j[\config'(\switch_j)]$ for some $i \ne j$, then $u \in V_i$ and $v \in V_j$. By construction, $uv \in E(G[\switch])$, so we configure $\switch$ such that $uv \in \switch[\config(\switch)]$. Replacing the subpath $u \to w \to v$ in $P$ by the edge $uv$ yields a path from $a$ to $b$ in $\x \sgraph \config$.

        If instead both $uw$ and $wv$ belong to $\switch_i[\config'(\switch_i)]$ for some $i < k$, then both $u$ and $v$ lie in $V_{i,j}$ with $j \coloneqq \config'(\switch_i)$. Since $G[\switch][V_{i,j}]$ is a clique, we have $uv \in E(G[\switch])$, so we set $\config(\switch)$ such that $uv \in \switch[\config(\switch)]$. Replacing $u \to w \to v$ with $uv$ in $P$ again yields a path from $a$ to $b$ in $\x \sgraph \config$.\qedhere
    \end{description}
\end{proof}

The complexity results of Szeider~\cite{Szeider03} for \tcp{}, together with the reductions and the transformation above, yield a complexity dichotomy for \lcs{} restricted to $\{1^\star\}$-bounded switches, depending on the shape of the switches. We establish its two sides separately.
Szeider showed that \tcp{} can be solved in linear time if every transition graph is complete multipartite~\cite[Proposition 9]{Szeider03}. Since \cref{con:switch-to-transition-graph} maps each switch to a transition graph isomorphic to its switch-induced graph, and all other transition graphs are complete, this result carries over to switch graphs:

\begin{lemma}
    \label{lem:dichotomy-poly}
    Let $\Pi \subseteq \mathcal{K}_\mathrm{multi}$ be a family of graphs without isolated vertices. Then \lcs{} restricted to $\{1^\star\}$-bounded $\Pi$-shaped switches can be solved in time $\bigO(n + m)$.
\end{lemma}
\begin{proof}
Fix a switch graph $\sgraph$ with $\{1^\star\}$-bounded $\Pi$-shaped switches. We construct the corresponding graph $G$ with transition system $T$ according to \cref{con:switch-to-transition-graph}. As discussed previously, the correctness of this reduction is guaranteed by \cref{lem:switch-to-transition-graph}. 

We begin by analyzing the running time of \cref{con:switch-to-transition-graph}. Creating a vertex in $G$ for each vertex and each switch in $\sgraph$ takes $\mathcal{O}(n + \sigma)$ time. For each switch $\switch \in \switches(\sgraph)$, we add $|V(G[\switch])|$ edges to $G$ and construct a transition graph for its corresponding switch vertex. This transition graph contains $|V(G[\switch])|$ vertices and $|E(G[\switch])|$ edges. Per switch, this amounts to $\mathcal{O}(|V(G[\switch])| + (|V(G[\switch])| + |E(G[\switch])|))$ time. Since $G[\switch]$ has no isolated vertices, we have $|V(G[\switch])| \le 2|E(G[\switch])|$, and thus the total per-switch running time simplifies to $\mathcal{O}(|E(G[\switch])|)$. Note that the transition graphs for the original vertices need not be stored explicitly and therefore do not affect the overall complexity. Thus, the total running time of the construction becomes
\[
    \mathcal{O}\left(n + \sigma + \sum_{\switch \in \switches(\sgraph)} |E(G[\switch])|\right) = \mathcal{O}(n + \sigma + m).
\]

Graph $G$ itself consists of $n + \sigma$ vertices and $|V(G[\switch])|$ edges for each switch $\switch \in \switches(\sgraph)$. Since $|V(G[\switch])| \le 2 |E(G[\switch])|$, this amounts to at most $\sum_{\switch \in \switches(\sgraph)} 2 |E(G[\switch])|$ edges in $G$. Deciding whether there exists a $T$-compatible path from $a$ to $b$ in $G$ can then be done in time $\mathcal{O}(|V(G)| + |E(G)|)$ using the algorithm of \cite[Proposition 9]{Szeider03}. This then yields a total running time of
\[
    \mathcal{O}(|V(G)| + |E(G)|) = \mathcal{O}\left(n + \sigma + \sum_{\switch \in \switches(\sgraph)} 2|E(G[\switch])|\right) = \mathcal{O}(n + \sigma + m).
\]

Thus, both the reduction and the algorithm for \tcp{} run in the same time, so the total running time amounts to $\mathcal{O}(n + \sigma + m)$. We can safely assume that each switch contains at least one edge, so $\sigma \le m$ (switches with no edges are not useful for solving the problem and can be configured arbitrarily). Thus, we can simplify the running time to $\mathcal{O}(n + m)$.
\end{proof}

Conversely, for every switch shape that is not complete multipartite, we obtain \np-hardness by reducing from \tcp{} via \cref{con:transition-graph-to-switch} and, where necessary, \cref{con:1-switch-to-star-switches}:

\begin{lemma}
    \label{lem:dichotomy-hard}
    Let $H$ be a graph without isolated vertices that is not complete multipartite. Then \lcs{} restricted to $\{1^\star\}$-bounded $H$-shaped switches is \np-hard.
\end{lemma}
\begin{proof}
We first characterize $H$ in a manner similar to the approach taken by \cite{Szeider03}. Since $H$ is not complete multipartite, it must contain vertices $u,v,w \in V(H)$ such that $uv \in E(H)$ but $uw, vw \notin E(H)$. Because $H$ has no isolated vertices, $w$ must be adjacent to some other vertex $x \in V(H)$. Depending on the adjacency between $x$ and $\{u,v\}$, the vertex-induced subgraph $H[\{u,v,w,x\}]$ is isomorphic to one of the following:
\begin{itemize}
    \item $K_2 + K_2$ if $x$ is not adjacent to $u$ or $v$,
    \item $P_4$ if $x$ is adjacent to exactly one of $\{u,v\}$,
    \item $L_{3,1}$ if $x$ is adjacent to both $u$ and $v$.
\end{itemize}

We now analyze these cases separately.
\begin{description}
    \item[$P_4$ or $L_{3,1}$ as a vertex-induced subgraph of $H$:]  
    As a consequence of \cite[Theorem~1]{Szeider03}, the problem \tcp{} is \np-complete even if every transition graph is isomorphic to $H$. Given such an instance $(G,T)$, we construct a switch graph $\sgraph_T$ restricted to $\{1^\star\}$-bounded switches using \cref{con:transition-graph-to-switch}, which is a correct reduction to \lcs{} by \cref{lem:transition-graph-to-switch}. The resulting switches are all $\{H,K_2\}$-shaped, with the $K_2$-shaped switches being static-edge switches.

    To eliminate these $K_2$-shaped switches, we replace each static-edge switch $\switch_e$ with a $\{1^\star\}$-bounded switch $\switch_e'$ whose switch-induced graph is isomorphic to $H$, containing $e$ in one of its positions. Except for the two endpoints of $e$, the vertex set consists of fresh vertices unique to $\switch_e'$. Since only the position containing $e$ can help towards connecting $a$ and $b$, $\switch_e'$ behaves exactly like the original static-edge switch. As a result, all switches are now $H$-shaped, and \lcs{} is \np-hard in this setting.

    \item[$K_2 + K_2$ as a vertex-induced subgraph of $H$:] Despite the fact that the problem \tcp{} is \emph{not} \np-complete when all transition graphs are isomorphic to $H$, it \emph{is} \np-complete when the transition graphs are restricted to $\{H,K_3\}$ \cite[Theorem~1]{Szeider03}. Reducing such an instance $(G,T)$ to a switch graph $\sgraph_T$ via \cref{con:transition-graph-to-switch} (as in the previous case) results in $\{1^\star\}$-bounded $\{H,K_3, K_2\}$-shaped switches.

    Applying \cref{con:1-switch-to-star-switches} to each $K_3$-shaped switch transforms it into three static-edge switches (since $K_3$ is complete tripartite; see \cref{fig:1-switch-to-star-switches}, left), which is a correct reduction by \cref{lem:1-switch-to-star-switches}. We then replace all static-edge switches as in the previous case, thus obtaining a switch graph restricted to $H$-shaped switches. Hence, \lcs{} is \np-hard in this setting as well.\qedhere
\end{description}
\end{proof}

Together, the two lemmas yield the following dichotomy.

\begin{theorem}
    \label{thm:dichotomy-lcs}
    For a family of graphs $\Pi$ without isolated vertices, \lcs{} restricted to $\{1^\star\}$-bounded $\Pi$-shaped switches is solvable in time $\bigO(n + m)$ if $\Pi \subseteq \mathcal{K}_\mathrm{multi}$; otherwise it is \np-complete.
\end{theorem}
\begin{proof}
If $\Pi \subseteq \mathcal{K}_\mathrm{multi}$, the claim follows from \cref{lem:dichotomy-poly}. Otherwise, $\Pi$ contains a graph $H$ that is not complete multipartite. Every switch graph whose switches are all $H$-shaped is in particular a switch graph with $\Pi$-shaped switches, so \cref{lem:dichotomy-hard} implies \np-hardness.
\end{proof}

Going beyond $\{1^\star\}$-bounded switches, we quickly encounter \np-hardness, even when restricting the switch shape to star graphs, as shown in the following theorem:

\begin{theorem}
    \label{thm:hardness-lcs-star-switches}
    \lcs{} is \np-complete, even when restricted to $\{2,1\}$-bounded $\mathcal{K}_{1,\star}$-shaped switches.
\end{theorem}
\begin{proof}
\tcp{} remains \np-complete even when each transition graph is isomorphic to the lollipop graph $L_{3,1}$ \cite{Szeider03}. Applying \cref{con:transition-graph-to-switch} to such an instance yields a switch graph restricted to $\{1^\star\}$-bounded $\{L_{3,1}, K_2\}$-shaped switches.

\cref{fig:1-switch-to-star-switches} (right) illustrates the application of \cref{con:1-switch-to-star-switches} to a $\{1^\star\}$-bounded $L_{3,1}$-shaped switch \switch{}. Let $V(G[\switch]) = \{a,b,c,d\}$ and $E(G[\switch])=\{\{a,b\},\allowbreak \{a,c\},\allowbreak \{b,c\},\allowbreak \{b,d\}\}$. We partition $V(G[\switch])$ into two sets: $V_0 = \{a,c,d\}$ and $V_1 = \{b\}$, such that all edges between the sets are contained in $E(G[\switch])$. The edge clique cover of $G[\switch][V_0]$ consists of the subsets $\{a,c\}$ and $\{d\}$, and that of $G[\switch][V_1]$ consists of $\{b\}$. This results in two star switches $\switch_0$ and $\switch_1$ with $\switch_0[0] = \{aw, cw\}$, $\switch_0[1] = \{dw\}$ and $\switch_1[0] = \{bw\}$, both of which are $\{2,1\}$-bounded $\mathcal{K}_{1,\star}$-shaped switches.

Hence, by applying \cref{con:1-switch-to-star-switches} to each switch, we obtain a switch graph where every switch is a $\{2,1\}$-bounded $\mathcal{K}_{1,\star}$-shaped switch. Since the original instance of \tcp{} is \np-complete and we have established the correctness of both the reduction to \lcs{} (\cref{lem:transition-graph-to-switch}) and the switch transformation (\cref{lem:1-switch-to-star-switches}), we conclude that \lcs{} restricted to $\{2,1\}$-bounded $\mathcal{K}_{1,\star}$-shaped switches is \np-hard.
\end{proof}

\subsection{Assigning Weights to Switch Positions}

We now turn to the parameterized complexity of a weighted variant of \lcs{}, formally defined as follows:

\begin{definition}
    \probdef{\wlcs{}}{
        A switch graph $\sgraph$, two distinct vertices $a,b \in V(\sgraph)$, a weight function $w: \switches(\sgraph) \times \integers \to \reals$ on switch positions, and a real $\lambda \in \reals$.
    }{
        Is there a configuration $\config \in \configs(\sgraph)$ such that $a$ and $b$ are connected in $\x{\sgraph}{\config}$ and $w(\config) \coloneqq \sum_{\switch \in \switches(\sgraph)} w(\switch, \config(\switch)) \le \lambda$?
    }
\end{definition}

Note that the problem remains \np-hard for constant $\lambda$ under the same structural restrictions as the unweighted variant, provided arbitrary weights are allowed. To analyze the parameterized complexity with respect to $\lambda$ more closely, we therefore consider a restricted variant with on-off switches and a fixed weight function $w(\switch,i) \coloneqq i$, called \oolcs{}, where the total weight of a configuration equals the number of switches set to the on-position.

In particular, we show that the problem is unlikely to be \fpt{} with respect to $\lambda$, even for very restricted switches. To this end, we provide a parameterized reduction from the \emph{weighted satisfiability problem} (\wsat{}), where we are given a propositional formula $\varphi$ in negation normal form (NNF)\footnote{A Boolean formula is in NNF if it contains only conjunctions, disjunctions, and negation operators that are directly applied to variables.} and a positive integer~$k \in \naturals$ and ask whether there exists a satisfying truth assignment~$I\colon \mathcal{V}(\varphi) \to \{0,1\}$ such that at most $k$ variables are set to true ($1$). More details on \wsat{} can be found in \cref{sec:wsat}.

In this reduction, we inductively construct a switch graph from the formula $\varphi$ based on its syntactic structure, in polynomial time. The resulting switch graph can be viewed as a multigraph obtained through \emph{series} and \emph{parallel compositions} of edges, where each variable is represented by an on-off switch.
Herein, we equip graphs with two ``endpoint'' vertices and, in a series composition, two graphs are joined by identifying one endpoint of each; in a parallel composition, both pairs of corresponding endpoints are identified.
This construction, illustrated in \cref{fig:sat-to-lcs} by an example, is formally defined as follows:

\begin{figure}[t]
    \centering
    \includegraphics[page=6]{graphics/figures.pdf}
    \caption{\cref{con:sat-to-switch-st} on $\varphi \coloneqq ((\textcolor{dodgerblue3}{x} \land \textcolor{chartreuse3}{\neg y}) \lor \textcolor{dodgerblue3}{\neg x} \lor (\textcolor{chartreuse3}{\neg y} \land  \textcolor{deeppink3}{z})) \land ((\textcolor{dodgerblue3}{x} \land \textcolor{chartreuse3}{y}) \lor \textcolor{deeppink3}{\neg z}) \land (\textcolor{deeppink3}{z} \lor \textcolor{dodgerblue3}{x} )$.}
    \label{fig:sat-to-lcs}
\end{figure}

\begin{construction}
\label{con:sat-to-switch-st}
    Given a formula $\varphi$ in NNF, we construct a corresponding switch graph $\sgraph_\varphi$, with an on-off switch $\switch_x$ for each variable $x \in \mathcal{V}(\varphi)$, as follows:
    \begin{itemize}
        \item If $\varphi \coloneqq \neg x$ for some variable $x$, then $\sgraph_\varphi$ consists of a single edge associated with the off-position of $\switch_x$.
        \item If $\varphi \coloneqq x$ for some variable $x$, then $\sgraph_\varphi$ consists of a single edge associated with the on-position of $\switch_x$.
        \item If $\varphi \coloneqq \psi \land \omega$, then $\sgraph_\varphi$ is obtained by the series-composition of $\sgraph_{\psi}$ and $\sgraph_{\omega}$.
        \item If $\varphi \coloneqq \psi \lor \omega$, then $\sgraph_\varphi$ is obtained by the parallel-composition of $\sgraph_{\psi}$ and $\sgraph_{\omega}$.
    \end{itemize}
\end{construction}

Suppose each switch $\switch_x$ is set to the on-position in a configuration $\config \in \configs(\sgraph_\varphi)$ exactly when the corresponding variable $x \in \mathcal{V}(\varphi)$ is set to true in a truth assignment $I$, implying $w(\config) = w(I)$. Intuitively, a literal is satisfied precisely when two endpoints of its corresponding switch graph are connected. Conjunctions are satisfied if and only if both subformulas evaluate to true---equivalently, the configured switch graph admits a path between its endpoints if and only if both subgraphs do. The same reasoning applies to disjunctions. Hence, $\x{\sgraph_\varphi}{\config}$ admits a path between its endpoints if and only if $\varphi$ is satisfied by $I$. Furthermore, if all variables in $\varphi$ are \emph{$(i,j)$-bounded}---that is, each variable occurs at most $i$ times in one polarity and at most $j$ times in the other---then every switch in $\sgraph_\varphi$ is also $\{i,j\}$-bounded. This yields the following lemma:

\begin{lemma}
    \label{lem:sat-to-switch-st}
    There is a parameter-linear reduction from \wsat{} restricted to $(i,j)$-bounded variables, parameterized by $k$, to \oolcs{} restricted to $\{i,j\}$-bounded on-off switches, parameterized by $\lambda$.
\end{lemma}
\begin{proof}
Let $(\varphi, k)$ be an instance of \wsat{}, and let $\sgraph_\varphi$ be the switch graph constructed in polynomial time from $\varphi$ as described in \cref{con:sat-to-switch-st}.
We prove that there exists a truth assignment $I$ with $I(\varphi) = 1$ and $w(I) \le k$ if and only if there exists a configuration $\config \in \configs(\sgraph_\varphi)$ such that the two endpoints are connected in $\x{\sgraph_\varphi}{\config}$ and $w(\config) \le \lambda = k$, by induction on the structure of $\varphi$.

Suppose $\config(\switch_x) = I(x)$ for all $x \in \mathcal{V}(\varphi)$. Then, $w(\config) = w(I) \le k = \lambda $. If $\varphi \coloneqq \neg x$ or $\varphi \coloneqq x$, then $I(\varphi) = 1$ if and only if the two endpoints are connected in $\x{\sgraph_\varphi}{\config}$, by construction. If $\varphi \coloneqq \psi \land \omega$, then $I(\varphi) = 1$ if and only if both $I(\psi) = 1$ and $I(\omega) = 1$. By induction, this holds if and only if there is a path between the endpoints in both $\x{\sgraph_{\psi}}{\config}$ and $\x{\sgraph_{\omega}}{\config}$, which is equivalent to a path in $\x{\sgraph_\varphi}{\config}$ due to the series composition. If $\varphi \coloneqq \psi \lor \omega$, then $I(\varphi) = 1$ if and only if at least one of $I(\psi) = 1$ or $I(\omega) = 1$. By induction, this holds if and only if there is a path between the endpoints in at least one of $\x{\sgraph_{\psi}}{\config}$ or $\x{\sgraph_{\omega}}{\config}$, which is equivalent to a path in $\x{\sgraph_\varphi}{\config}$ due to the parallel composition.
\end{proof}

The class \w{SAT} consists of all parameterized problems that admit a parameterized reduction to \wsat{} on arbitrary propositional formulas~\cite[Definition~7.37]{Flum06}. Hence, the reduction of \cref{lem:sat-to-switch-st} on unbounded formulas proves the following:

\begin{theorem}
    \label{thm:hardness-wstcon}
    \oolcs{} is \w{SAT}-hard parameterized by~$\lambda$.
\end{theorem}
Note that $\w{t} \subseteq \w{SAT}$ for all~$t \ge 1$. Therefore, \cref{thm:hardness-wstcon} implies that \oolcs{} for unrestricted switches is unlikely to be \fpt{} with respect to $\lambda$.

To investigate the parameterized complexity with bounded switches, we reduce from \wsat{} variants with bounded variables for which we can establish \wone{}-hardness. For instance, in \cref{sec:wsat} we give a parameter-linear reduction from \psgi{} to \wsat{} restricted to $(1,1)$-bounded variables (\cref{thm:psi-to-wsat-11}). Thus, we obtain the following for \oolcs{}:
\begin{theorem}
    \label{thm:wlcs-11-switches}
    \oolcs{} restricted to $\{1,1\}$-bounded switches is \wone-hard parameterized by~$\lambda$. Moreover, there is a parameter-linear reduction from \psgi{} to it.
\end{theorem}
\begin{proof}
    Composing the parameter-linear reduction of \cref{thm:psi-to-wsat-11} with the one from \cref{lem:sat-to-switch-st} (with $i = j = 1$) yields a parameter-linear reduction from \psgi{} to \oolcs{} restricted to $\{1,1\}$-bounded switches. Since \psgi{} is \wone-hard, so is \oolcs{} restricted to $\{1,1\}$-bounded switches.
\end{proof}

\begin{figure}[t]
    \centering
    \includegraphics[page=14]{graphics/figures.pdf}
    \caption{Transforming a $\{1,1\}$-bounded switch into four $\{2,1\}$-bounded $\mathcal{K}_{1,\star}$-shaped switches, allowing for the same connections between original vertices.}
    \label{fig:transform-11-to-21-switches}
\end{figure}

This result carries over to $\{2,1\}$-bounded $\mathcal{K}_{1,\star}$-shaped switches by replacing each $\{1,1\}$-bounded switch by four such switches, as described in \cref{con:transform-11-to-21-switches} and illustrated in \cref{fig:transform-11-to-21-switches}.%

\begin{construction}
    \label{con:transform-11-to-21-switches}
    Let $\sgraph$ be a switch graph and $\switch \in \switches(\sgraph)$ a $\{1,1\}$-bounded switch with $\switch[0] = \{v_0v_2\}$ and $\switch[1] = \{v_1v_3\}$. We construct a new switch graph $\sgraph'$ from \sgraph{} by introducing six fresh vertices $\{u_0,u_1,u_2,u_3,w_0,w_1\}$ and replacing $\switch$ with four new $\{2,1\}$-bounded $\mathcal{K}_{1,\star}$-shaped switches $\{\switch_0, \switch_1, \switch_2, \switch_3\}$, such that for each $i,j \in \{0,1\}$, we have:
    \[
        \switch_{i+2j}[i] \coloneqq \{v_{i+2j}u_{i+2j}\} \quad \text{ and }\quad \switch_{i+2j}[1-i] \coloneqq \{u_{2j}u_{1+2j}, u_{i+2j}w_{i}\}.
    \]
\end{construction}
Consider one transformed $\{1,1\}$-bounded switch \switch. Intuitively, setting all four corresponding switches in the transformed switch graph to position~$0$ realizes the path $\langle v_0, u_0, u_1, w_1, u_3, u_2, v_2 \rangle$, which takes the place of edge $v_0v_2 \in \switch[0]$, while setting all of them to position~$1$ realizes the path $\langle v_1, u_1, u_0, w_0, u_2, u_3, v_3 \rangle$, which takes the place of edge $v_1v_3 \in \switch[1]$. Conversely, any path passing through the fresh vertices of the transformed switch can only follow one of these two routes, which requires all four switches to be set to the same position. Since turning $\switch$ on thus corresponds to turning all four new switches on, the weight of a configuration scales by a factor of four, which keeps the reduction parameter-linear.
\begin{theorem}
    \label{thm:wlcs-star-switches}
    \oolcs{} restricted to $\{2,1\}$-bounded $\mathcal{K}_{1,\star}$-shaped switches is \wone-hard parameterized by $\lambda$. Moreover, there is a parameter-linear reduction from \psgi{} to it.
\end{theorem}
\begin{proof}
Let $\sgraph$ be a switch graph consisting of $\{1,1\}$-bounded switches and $\lambda$ the maximum solution weight. Without loss of generality, we may assume that $\sgraph$ contains no static-edge switches, as we can always add a second empty position to those switches. Moreover, we may assume that each switch position contains at least one edge. Otherwise, we introduce two fresh vertices $u$ and $v$, and insert the edge $uv$ into all empty positions. Since $u$ and $v$ are isolated from the rest of the graph, this modification does not affect the connectivity between the original vertices.

To obtain a valid parameter-linear reduction from \oolcs{}, parameterized by $\lambda$ and restricted to $\{1,1\}$-bounded switches, to the same parameterized problem restricted to $\{2,1\}$-bounded $\mathcal{K}_{1,\star}$-shaped switches, we apply \cref{con:transform-11-to-21-switches} to every switch, resulting in a new switch graph $\sgraph'$. Furthermore, we set the new maximum weight to $\lambda'\coloneqq 4\lambda$. Clearly, the reduction can be achieved in polynomial time, and $\lambda' \in \bigO(\lambda)$.

To show correctness of the reduction, we aim to show that there exists a configuration $\config \in \configs(\sgraph)$ such that $\x {\sgraph} \config$ admits a path from $a$ to $b$ and $w(\config) \leq \lambda$ if and only if there exists a configuration $\config' \in \configs(\sgraph')$ such that $\x {\sgraph'} \config'$ admits a path from $a$ to $b$ and $w(\config') \leq \lambda'$.

\begin{description}
    \item[$(\Rightarrow)$] Let $\config$ be a configuration of $\sgraph$ such that $\x {\sgraph} \config$ admits a path from $a$ to $b$ and $w(\config) \leq \lambda$. We now construct a configuration $\config' \in \configs(\sgraph')$ such that, for each switch $\switch \in \switches(\sgraph)$ and $i < 4$, we have $\config'(\switch_i) \coloneqq \config(\switch)$. Clearly, $w(\config') = 4 w(\config) \leq 4\lambda = \lambda'$. What remains to be shown is that $\x {\sgraph'} {\config'}$ admits a path from $a$ to $b$.
    
    Let $e$ be an arbitrary edge in $\x{\sgraph}{\config}$, and let $\switch \in \switches(\sgraph)$ be the switch such that $e \in \switch[i]$ for $i \coloneqq \config(\switch)$. Denote the endpoints of $e$ as $v_{i}$ and $v_{i+2}$. For each $j \in \{0,1\}$, we have $v_{i+2j}u_{i+2j} \in \switch_{i+2j}[i]$ and $u_{2j}u_{1+2j}, u_{(1-i)+2j}w_{1-i} \in \switch_{(1-i)+2j}[i]$, resulting in a path from $v_{i+2j}$ to $w_{1-i}$ in $\x{\sgraph'}{\config'}$, where $w_{1-i}$ and all other inner vertices of the path are auxiliary vertices introduced by the transformation of switch $\switch$. Combining these paths for $j=0$ and $j=1$ yields a path from $v_{i}$ to $v_{i+2}$.
    
    Consequently, each edge $uv$ in the path from $a$ to $b$ in $\x {\sgraph} \config$ can be replaced by a corresponding path between $u$ and $v$ in $\x {\sgraph'} {\config'}$, thus resulting in a path from $a$ to $b$ in $\x {\sgraph'} {\config'}$.

    \item[$(\Leftarrow)$] Let $\config'$ be a configuration of $\sgraph'$ such that $\x {\sgraph'} \config'$ admits a path $P'$ from $a$ to $b$ and $w(\config') \leq \lambda'$. We construct a configuration $\config \in \configs(\sgraph)$ such that, for each switch $\switch \in \switches(\sgraph)$, we have $\config(\switch) \coloneqq \min_{i < 4} \{\config'(\switch_i)\}$. Thus, $\config(\switch) \le \frac{1}{4} \sum_{i < 4} \config'(\switch_i)$, and consequently $w(\config) \leq \frac{1}{4} w(\config') \leq \frac{1}{4} \lambda' = \lambda$. It remains to show that $\x {\sgraph} \config$ admits a path from $a$ to $b$.
    
    To do so, we process the vertices in $P'$ sequentially and construct a path $P$ from $a$ to $b$ in $\x {\sgraph} \config$ in parallel. The path $P'$ starts and ends with original vertices of $\sgraph$ (specifically, $a$ and $b$), so any auxiliary vertices introduced by the transformation must be internal and followed by further vertices.
    
    Initially, let $v$ be the first vertex in $P'$, and initialize $P \coloneqq \langle v \rangle$.

    Assume $v$ is not the last vertex in $P'$. Then, by construction, it must be followed by an auxiliary vertex introduced by the transformation of some $\{1,1\}$-bounded switch $\switch$ in $\sgraph$. 
    In particular, there exists a switch $\switch_{i+2j}$ with $i = \config'(\switch_{i+2j})$ and $j \in \{0,1\}$ such that $\switch_{i+2j}[i] = \{v_{i+2j}u_{i+2j}\}$, where $v = v_{i+2j}$ and $u_{i+2j}$ is the next vertex following $v$ in $P'$.

    The only other switch incident to $u_{i+2j}$ is $\switch_{(1-i)+2j}$ via an edge $u_{2j}u_{1+2j}$ contained in position $i$. Thus, we have $\config'(\switch_{(1-i)+2j}) = i$ and $u_{i+2j}$ is followed by the auxiliary vertex $u_{(1-i)+2j}$ in $P'$.

    Next, since $u_{(1-i)+2j}$ is not incident to any switches other than $\switch_{(1-i)+2j}$, and $\switch_{(1-i)+2j}[i]$ contains exactly one additional edge $u_{(1-i)+2j}w_{1-i}$, it follows that $u_{(1-i)+2j}$ is followed by the auxiliary vertex $w_{1-i}$ in $P'$.

    The vertex $w_{1-i}$ has only one additional incident switch: $\switch_{(1-i)+2(1-j)}$ via the edge $w_{1-i}u_{(1-i)+2(1-j)}$ contained in position $i$. Hence, $\config'(\switch_{(1-i)+2(1-j)}) \allowbreak= i$ and $w_{1-i}$ is followed by the auxiliary vertex $u_{(1-i)+2(1-j)}$ in $P'$.
    
    Although $u_{(1-i)+2(1-j)}$ is incident to switch $\switch_{i+2(1-j)}$, it is through the \ifjournal same \fi edge $u_{2(1-j)}u_{1+2(1-j)}$ already contained in $\switch_{(1-i)+2(1-j)}[i]$. As there are no other switches or edges in $\switch_{(1-i)+2(1-j)}[i]$ incident to $u_{(1-i)+2(1-j)}$, this vertex must be followed by the auxiliary vertex $u_{i+2(1-j)}$ in $P'$.

    At this point, the only other switch incident to $u_{i+2(1-j)}$ is $\switch_{i+2(1-j)}$. Suppose this switch is configured to position $1-i$, consisting of edges $u_{2(1-j)}u_{1+2(1-j)}$ and $u_{i+2(1-j)}w_i$. Since $P'$ is a path, we cannot return to the previously visited vertex $u_{(1-i)+2(1-j)}$, so the only possible successor of $u_{i+2(1-j)}$ would be the auxiliary vertex $w_i$. However, $w_i$ is only incident to switch $\switch_{i+2j}$ via an edge in position $1-i$, while $\switch_{i+2j}$ is already configured to position $i$. This creates a contradiction, as the path cannot continue through $w_i$. Therefore, switch $\switch_{i+2(1-j)}$ must be configured to position $i$, which contains a single edge $u_{i+2(1-j)} v_{i+2(1-j)}$. Hence, $u_{i+2(1-j)}$ is followed by the original vertex $v_{i+2(1-j)}$ in $P'$.

    In summary, for all $\ell < 4$, we have $\config'(\switch_\ell) = i$, resulting in $\config(\switch) = i$. By construction, edge $v_{i+2j}v_{i+2(1-j)}$ is contained in position $i$ of switch $\switch$, and thus part of $\x {\sgraph} \config$. Since the last vertex added to $P$ was $v_{i+2j}$, we extend $P$ by appending $v_{i+2(1-j)}$. 
    
    We repeat the procedure for $v = v_{i+2(1-j)}$. Eventually, this constructs a full path $P$ from $a$ to $b$ in $\x {\sgraph} \config$.
\end{description}

Composing this parameter-linear reduction with the one from \psgi{} to \oolcs{} restricted to $\{1,1\}$-bounded switches given by \cref{thm:wlcs-11-switches} yields a parameter-linear reduction from \psgi{} to \oolcs{} restricted to $\{2,1\}$-bounded $\mathcal{K}_{1,\star}$-shaped switches. Since \psgi{} is \wone-hard, so is \oolcs{} restricted to $\{2,1\}$-bounded $\mathcal{K}_{1,\star}$-shaped switches.
\end{proof}

\section{Global Connectivity}
\label{sec:gcon}

We now study configurations that not only connect a pair of vertices but ensure that the entire graph is connected, formalized in the following decision problem:
\begin{definition}
    \probdef{\gcs{}}{
        A switch graph \sgraph{}.
    }{
        Is there a configuration $\config \in \configs(\sgraph)$ such that $\x{\sgraph}{\config}$ is connected?
    }
\end{definition}

The weighted variants \wgcs{} and \oogcs{} are defined analogously to their local connectivity counterparts in \cref{sec:stcon}.

\subsection{Formulation as Matroids}

We assume familiarity with basic matroid concepts and refer to Oxley~\cite{Oxley92} for an introduction to matroid theory.
Katz et al.\ showed that \gcs{} restricted to $\{1^\star\}$-bounded $\mathcal{K}_{1,\star}$-shaped switches can be solved in polynomial time via a reduction to \probname{Matroid Intersection}, which asks for a largest common independent set in two matroids over the same ground set~\cite{Katz12}.
Given a switch graph $\sgraph$ under these restrictions, they define two matroids on $E(\sgraph)$: a \emph{graphic matroid} $(E(\sgraph), \mathcal{I}_1)$, where a set of edges is independent if it is acyclic, and a \emph{partition matroid} $(E(\sgraph), \mathcal{I}_2)$, where a set of edges is independent if it contains at most one edge from each switch.
A connecting configuration exists if and only if there is a spanning tree realizable by a configuration, or equivalently, an independent set $A \in \mathcal{I}_1 \cap \mathcal{I}_2$ of size $n - 1$.
This reduces the problem to \probname{Matroid Intersection}, which can be solved in polynomial time~\cite{Chakrabarty19}.

Importantly, this argument does not depend on the switches being $\mathcal{K}_{1,\star}$-shaped, and thus generalizes to all $\{1^\star\}$-bounded switches. We can generalize even further by restricting attention to switches that admit a matroid structure, that is, switches $\switch \in \switches(\sgraph)$ for which $(E(G[\switch]), \mathcal{I}_{\switch})$ with $\mathcal{I}_{\switch} \coloneqq \{ A \subseteq E(G[\switch]) \mid \exists i < |\switch|\colon A \subseteq \switch[i] \}$ forms a matroid. We call $(E(G[\switch]), \mathcal{I}_{\switch})$ the \emph{switch matroid} associated with $\switch$, and we refer to such switches as \emph{matroid switches}. 
We then construct the matroid $(E(\sgraph), \mathcal{I}_3)$ as the direct sum of the switch matroids, so that $\mathcal{I}_3$ contains all edge sets that choose edges from at most one position per switch. As before, connecting configurations then correspond to common independent sets of size $n - 1$ in $\mathcal{I}_1$ and $\mathcal{I}_3$, which yields the following theorem:

\begin{theorem}
    \label{thm:gcs-matroid-switches}
    \gcs{} restricted to matroid switches can be decided in time $\bigO(n^2 m \log n \cdot T)$, where $T$ is the running time of the independence test of a switch matroid.
\end{theorem}
\begin{proof}
Given an instance $\sgraph$ of \gcs{}, we define matroids $(E(\sgraph), \mathcal{I}_1)$ and $(E(\sgraph), \mathcal{I}_3)$ as previously described. We first show that $\sgraph$ admits a connecting configuration if and only if there exists an independent set $A \in \mathcal{I}_1 \cap \mathcal{I}_3$ with $|A| = n - 1$.
\begin{description}
    \item[$(\Rightarrow)$] Let $\config$ be a configuration such that $\x{\sgraph}{\config}$ is connected, and let $T$ be a spanning tree of $\x{\sgraph}{\config}$. The edges of $T$ form our set $A$, where each $e \in A$ is associated with a switch~$\switch$ such that $e \in \switch[\config(\switch)]$. Then $|A| = n - 1$ since $T$ is a spanning tree. Moreover, $A \in \mathcal{I}_1$ since $T$ is acyclic, and $A \in \mathcal{I}_3$ since all edges of $A$ lie in the positions selected by~$\config$.

    \item[$(\Leftarrow)$] Let $A \in \mathcal{I}_1 \cap \mathcal{I}_3$ with $|A| = n - 1$. Since $A$ is acyclic and has $n - 1$ edges, it forms a spanning tree on $V(\sgraph)$. Since $A \in \mathcal{I}_3$, for each switch $\switch$, the edges of $A$ associated with $\switch$ are independent in its switch matroid, that is, they are contained in a single position $\switch[i]$, and we set $\config(\switch) \coloneqq i$. Then every edge of $A$ is contained in $\x{\sgraph}{\config}$, so $\x{\sgraph}{\config}$ contains a spanning tree and is thus connected.
\end{description}
Since every independent set in $\mathcal{I}_1$ has at most $n - 1$ elements, it suffices to compute a largest common independent set and check whether it has size $n - 1$.

The matroid intersection problem can be solved in time $\bigO(m'r\log rT_\mathrm{ind})$, where $m'$ is the size of the ground set, $r$ is the largest possible cardinality, and $T_\mathrm{ind}$ is the maximum running time of the independence test of either matroid \cite{Chakrabarty19}. In our case, we have $r \le n - 1$ and $m' = m$.

Since we have at most $r$ edges in the independent set, testing whether it contains a cycle can be achieved in $\bigO(r)$ time using depth first search. Testing whether all edges are from at most one position per switch, we split up the independent set into its switch components and test if each component is independent in its respective switch matroid. Splitting up the edges requires $\bigO(r)$ time, and as we have at most $r$ switches, the entire independence test for $(E(\sgraph), \mathcal{I}_3)$ takes $\bigO(r + rT)$ time. As a result, we have $T_\mathrm{ind} = \bigO(r \cdot T)$.

Since the matroids are defined implicitly, the reduction does not incur additional overhead. The overall complexity of solving the matroid intersection instance becomes:
\[
    \bigO(m'r\log r \cdot T_\mathrm{ind}) = \bigO(m' r^2 \log r \cdot T) = \bigO(n^2 m \log n \cdot T). \qedhere
\]
\end{proof}

Observe that $\{1^\star\}$-bounded switches are a special case of matroid switches, where each switch matroid is a uniform matroid of rank one, allowing independence to be checked in constant time. This immediately yields the following corollary:

\begin{corollary}
    \gcs{} restricted to $\{1^\star\}$-bounded switches can be decided in $\bigO(n^2 m \log n)$ time.
\end{corollary}

Finally, to handle the weighted case, we leverage a weighted variant of \probname{Matroid Intersection}, defined as follows:
\begin{definition}
    \prob{\probname{Weighted Exact Matroid Intersection}}{Input:}{
        Two matroids $(E, \mathcal{I}_1)$ and $(E, \mathcal{I}_2)$ over the same ground set $E$, a weight function $w \colon E \to \reals$, and a positive integer $k \in \naturals$.
    }{Task:}{
Find an independent set $A \in \mathcal{I}_1 \cap \mathcal{I}_2$ of size exactly $k$ minimizing $w(A) = \sum_{e \in A} w(e)$ or correctly report that none exists.
    }
\end{definition}

Given an instance $(\sgraph, w, \lambda)$ of \wgcs{} restricted to $\{1^\star\}$-bounded switches, we first assign each edge $e \in E(G[\switch])$ of each switch $\switch \in \switches(\sgraph)$ the minimum weight over all switch positions containing~$e$, that is, $w(e) \coloneqq \min_{i < |\switch|} \{ w(\switch,i) \mid e \in \switch[i]\}$.
Finding a minimum spanning tree realizable by a configuration then corresponds to finding a minimum-weight independent set $A \in \mathcal{I}_1 \cap \mathcal{I}_2$ of size $n - 1$.

However, the weight of the spanning tree does not necessarily equal the weight of the corresponding configuration, as switches that contribute no edge to the spanning tree may still incur additional weight in the configuration. To address this, we shift edge weights as follows: For each switch $\switch \in \switches(\sgraph)$ and each edge~$e \in E(G[\switch])$, define $w'(e) \coloneqq w(e) - \min_{i < |\switch|} w(\switch,i)$, and let $W \coloneqq \sum_{\switch \in \switches(\sgraph)} \min_{i < |\switch|} w(\switch,i)$. Observe that every connecting configuration~$\config$ satisfies $w(\config) \ge W + w'(T)$ for each spanning tree~$T$ of $\x{\sgraph}{\config}$, with equality if $\config$ is a cheapest configuration realizing~$T$. Thus, connecting configurations of weight at most $\lambda$ correspond to common independent sets of size $n - 1$ and weight at most $\lambda - W$ with respect to $w'$, which can be found using \probname{Weighted Exact Matroid Intersection} in polynomial time~\cite{Brezovec86}. As a result, we obtain the following theorem:

\begin{theorem}
    \label{thm:wgcs-matroid}
    \wgcs{} restricted to $\{1^\star\}$-bounded switches can be decided in $\bigO(nm \cdot (n + \log m))$ time.
\end{theorem}
\begin{proof}
Given an instance $(\sgraph, w, \lambda)$ of \wgcs{}, we define matroids $(E(\sgraph), \mathcal{I}_1)$ and $(E(\sgraph), \mathcal{I}_2)$ and the weight function $w'$ as previously described. We show that $\sgraph$ admits a connecting configuration of weight at most $\lambda$ if and only if there exists a set $A \in \mathcal{I}_1 \cap \mathcal{I}_2$ with $|A| = n - 1$ and $w'(A) \le \lambda - W$.
\begin{description}
    \item[$(\Rightarrow)$] Let $\config$ be a connecting configuration with $w(\config) \le \lambda$, and let $T$ be a spanning tree of $\x{\sgraph}{\config}$. The edges of $T$ form our set $A$, where each $e \in A$ is associated with a switch~$\switch$ such that $\{e\} = \switch[\config(\switch)]$. In particular, distinct edges of $A$ are associated with distinct switches, so $A \in \mathcal{I}_2$. Moreover, $|A| = n - 1$ and $A \in \mathcal{I}_1$, since $T$ is a spanning tree. Let $\switches_A \subseteq \switches(\sgraph)$ denote the switches associated with edges of $A$. For each $e \in A$ and its corresponding switch $\switch \in \switches_A$, we have $w'(e) = w(e) - \min_{i < |\switch|} w(\switch,i) \le w(\switch, \config(\switch)) - \min_{i < |\switch|} w(\switch,i)$. Hence,
    \begin{align*}
        w'(A) = \sum_{e \in A} w'(e) &\le \sum_{\switch \in \switches_A} \Big(w(\switch,\config(\switch)) - \min_{i<|\switch|} w(\switch,i)\Big)\\
        &\le \sum_{\switch \in \switches(\sgraph)} \Big(w(\switch,\config(\switch)) - \min_{i<|\switch|} w(\switch,i)\Big) \journalbreak= w(\config) - W \le \lambda - W.
    \end{align*}

    \item[$(\Leftarrow)$] Let $A \in \mathcal{I}_1 \cap \mathcal{I}_2$ with $|A| = n - 1$ and $w'(A) \le \lambda - W$. Since $A$ is acyclic and has $n - 1$ edges, it forms a spanning tree on $V(\sgraph)$. Since $A \in \mathcal{I}_2$, distinct edges of $A$ are associated with distinct switches. Let $\switches_A \subseteq \switches(\sgraph)$ denote these switches. For each $e \in A$ and its corresponding switch $\switch \in \switches_A$, we set $\config(\switch)$ to a cheapest position containing~$e$, so that $w(\switch, \config(\switch)) = w(e) = w'(e) + \min_{i < |\switch|} w(\switch,i)$. Every other switch $\switch \notin \switches_A$ is set to a cheapest position, so that $w(\switch, \config(\switch)) = \min_{i < |\switch|} w(\switch,i)$. Then $\x{\sgraph}{\config}$ contains all edges of $A$ and is thus connected. Moreover,
    \begin{align*}
        w(\config) &= \sum_{\switch \in \switches_A} w(\switch, \config(\switch)) + \sum_{\switch \in \switches(\sgraph) \setminus \switches_A} w(\switch, \config(\switch))\\
        &= \sum_{e \in A} w'(e) + \sum_{\switch \in \switches_A} \min_{i<|\switch|} w(\switch,i) + \sum_{\switch \in \switches(\sgraph) \setminus \switches_A} \min_{i<|\switch|} w(\switch,i)\\
        &= w'(A) + W \le \lambda.
    \end{align*}
\end{description}
Hence, it suffices to solve \probname{Weighted Exact Matroid Intersection} with $k \coloneqq n - 1$ and to check whether the returned set exists and has weight at most $\lambda - W$ with respect to $w'$.

The construction of the weight function $w'$ can be achieved in $\bigO(m)$ time as we can bound the number of switch positions by $m$. 

\probname{Weighted Exact Matroid Intersection} can be solved in time $\bigO(m'k(k+T_\mathrm{circuit} + \log m'))$, where $m'$ is the size of the ground set and $T_\mathrm{circuit}$ is the maximum running time of the circuit finding oracle of either matroid \cite{Brezovec86}. In our case, we have $k = n - 1$ and $m' = m$.

The circuit finding oracle, given an independent set $A \subseteq E(\sgraph)$ and an additional edge $e \in E(\sgraph) \setminus A$, must determine whether $A \cup \{e\}$ remains independent or else return a circuit contained in $A \cup \{e\}$. For the graphic matroid, this requires finding a cycle, which can be achieved in $\bigO(k)$ time using depth first search. For the partition matroid, we just need to identify if an edge in $A$ belongs to the same switch as $e$, which can be done in $\bigO(k)$ as well.

Thus, the overall running time amounts to:
\[
    \bigO(m) + \bigO(m'k(k+k+\log m')) = \bigO(nm \cdot (n + \log m)) . \qedhere
\]
\end{proof}

\subsection{Satisfiability Reductions}

So far, we have seen that \gcs{} and \wgcs{} can be solved in polynomial time for certain restricted switch classes. To complement these results, we now establish hardness by providing a construction that encodes a formula $\varphi$ in conjunctive normal form (CNF) as a switch graph. For ease of notation, we view $\varphi$ as a set of clauses, where each clause $C \in \varphi$ is a set of literals over $\mathcal{V}(\varphi)$. The construction of $\sgraph_\varphi$, illustrated in \cref{fig:sat-to-gcs} by an example, is defined as follows:

\begin{figure}[t]
    \centering
        \includegraphics[page=7]{graphics/figures.pdf}
        \caption{\cref{con:sat-to-gcs} on $\varphi \coloneqq \protect\underbrace{\left(\textcolor{dodgerblue3}{x}\right)}_{C_1} \land \protect\underbrace{\left(\textcolor{chartreuse3}{y} \lor \textcolor{deeppink3}{\neg z}\right)}_{C_2} \land \protect\underbrace{\left(\textcolor{dodgerblue3}{\neg x} \lor \textcolor{chartreuse3}{\neg y} \lor \textcolor{deeppink3}{z}\right)}_{C_3} \land \protect\underbrace{\left(\textcolor{dodgerblue3}{x} \lor \textcolor{deeppink3}{z}\right)}_{C_4}$.}
        \label{fig:sat-to-gcs}
\end{figure}

\begin{construction}
    \label{con:sat-to-gcs}
    Given a CNF formula $\varphi$, we construct a switch graph $\sgraph_\varphi$ with $V(\sgraph_\varphi) \coloneqq \{u\}\, \cup\, \{v_C \mid C \in \varphi\}$ and an on-off switch $\switch_x$ for each variable $x \in \mathcal{V}(\varphi)$, where $\switch_x[0] \coloneqq \{ uv_C \mid \neg x \in C \}$ and $\switch_x[1] \coloneqq \{ uv_C \mid x \in C \}$.
\end{construction}  

Suppose each switch $\switch_x$ is set to the on-position in a configuration $\config \in \configs(\sgraph_\varphi)$ if and only if the corresponding variable $x \in \mathcal{V}(\varphi)$ is set to true in a truth assignment $I$. Intuitively, for any clause $C \in \varphi$ and literal $\ell \in C$, the literal is satisfied exactly when the corresponding switch contributes an edge between $u$ and $v_C$. Hence, $\x {\sgraph_\varphi} \config$ is connected if and only if $\varphi$ is satisfied by $I$. Moreover, each switch is $\mathcal{K}_{1,\star}$-shaped, and if all variables in $\varphi$ are $(i,j)$-bounded, then every switch in $\sgraph_\varphi$ is $\{i,j\}$-bounded as well. This yields the following lemma:

\begin{lemma}
    \label{lem:sat-to-gcs}
    There is a polynomial-time many-one reduction from \sat{}, restricted to CNF formulas with $(i,j)$-bounded variables, to \gcs{}, restricted to $\{i,j\}$-bounded $\mathcal{K}_{1,\star}$-shaped on-off switches. Moreover, under the same restrictions, there is a parameter-linear reduction from \wsat{}, parameterized by $k$, to \oogcs{}, parameterized by $\lambda$.
\end{lemma}
\begin{proof}
Let $\varphi$ be a CNF formula with $(i,j)$-bounded variables, and let $\sgraph_\varphi$ be the respective switch graph obtained from \cref{con:sat-to-gcs}. To show that there exists a satisfying truth assignment $I$ of $\varphi$ if and only if there exists a configuration $\config \in \configs(\sgraph_\varphi)$ such that $\x {\sgraph_\varphi} \config$ is connected, suppose $\config(\switch_x) = I(x)$ for all $x \in \mathcal{V}(\varphi)$.

By definition, $I(\varphi) = 1$ if and only if $I(C) = 1$ for every clause $C \in \varphi$. For each clause $C \in \varphi$, this holds if and only if there is a variable $x \in \mathcal{V}(\varphi)$ such that $x \in C$ and $I(x) = 1$, or $\neg x \in C$ such that $I(x) = 0$. This is equivalent to the existence of a switch $\switch_x \in \switches(\sgraph_\varphi)$ such that $\config(\switch_x) = 0$ and $uv_C \in \switch_x[0]$, or $\config(\switch_x) = 1$ and $uv_C \in \switch_x[1]$. Hence, $I(C) = 1$ for all $C \in \varphi$ if and only if $uv_C \in E(\x {\sgraph_\varphi} \config)$ for all $C \in \varphi$, or equivalently, if $\x {\sgraph_\varphi} \config$ is connected. Therefore,~$\varphi$ is satisfiable if and only if there exists a connecting configuration of $\sgraph_\varphi$. 

This establishes the correctness of the reduction from \sat{} to \gcs{}. Furthermore, since $w(I) = w(\config)$ by construction, setting $\lambda \coloneqq k$ yields a parameter-linear reduction from \wsat{} to \oogcs{}.
\end{proof}

Since \sat{} remains \np{}-complete when restricted to CNF formulas with $(2,1)$-bounded variables~\cite{Tovey84}, the following is a direct consequence of \cref{lem:sat-to-gcs}:
\begin{theorem}
  \label{thm:gcs-21-npc}
    \gcs{} is \np{}-complete, even when restricted to $\{2,1\}$-bounded $\mathcal{K}_{1,\star}$-shaped switches.
\end{theorem}

Note that a $\{2,1\}$-bounded switch is the simplest switch that is not necessarily a matroid switch, and already at this point the problem becomes intractable.

Turning to the weighted variant, we again reduce from a bounded variant of \wsat{}. In \cref{sec:wsat}, we give a parameter-linear reduction from \psgi{} to \wsat{} restricted to CNF formulas with $(2,1)$-bounded variables (\cref{thm:wsat-gcs}). Thus, we obtain the following for \oogcs{}:
\begin{theorem}
    \label{thm:wgcs-star-switches}
    \oogcs{} restricted to $\{2,1\}$-bounded $\mathcal{K}_{1,\star}$-shaped switches is \wone-hard parameterized by~$\lambda$. Moreover, there is a parameter-linear reduction from \psgi{} to it.
\end{theorem}
\begin{proof}
    Composing the parameter-linear reduction of \cref{thm:wsat-gcs} with the one from \cref{lem:sat-to-gcs} (with~$i = 2$ and $j = 1$) yields a parameter-linear reduction from \psgi{} to \oogcs{} restricted to $\{2,1\}$-bounded $\mathcal{K}_{1,\star}$-shaped switches. Since \psgi{} is \wone-hard, so is \oogcs{} restricted to $\{2,1\}$-bounded $\mathcal{K}_{1,\star}$-shaped switches.
\end{proof}

\section{Weighted Satisfiability with Bounded Occurrences}
\label{sec:wsat}

In this section, we provide the parameter-linear reductions from \psgi{} to bounded variants of \wsat{} that complete the \wone-hardness proofs in \cref{sec:stcon,sec:gcon}. These reductions are also of independent interest. In particular, \cref{thm:psi-to-wsat-11} answers an open question by Kanj and Szeider~\cite{Kanj15} on the parameterized complexity of \wsat{} restricted to $3$-normalized formulas with at most two occurrences per variable.

Recall that \wsat{} is defined as follows:

\begin{definition}
    \probdef{\wsat{}}{
        A propositional formula $\varphi$ in NNF and a positive integer~$k \in \naturals$.
    }{
        Is there a truth assignment $I\colon \mathcal{V}(\varphi) \to \{0,1\}$ with $I(\varphi) = 1$ and $w(I) \coloneqq \sum\limits_{\mathclap{x \in \mathcal{V}(\varphi)}} I(x) \le k$?
    }
\end{definition}

Throughout, \wsat{} is parameterized by~$k$. We are interested in variants where each variable occurs only a bounded number of times. In particular, we call a variable \emph{$(i,j)$-bounded} if it occurs at most $i$ times in one polarity and at most $j$ times in the other.

Our starting point is a simple reduction from \psgi{} (\cref{def:psi}). Its variables encode a choice of vertices and edges in $G$, such that satisfying assignments of small weight correspond exactly to valid subgraph isomorphisms from $H$ to $G$. To address each edge in both directions, let $A(G) \coloneqq \{\overrightarrow{uv}, \overrightarrow{vu} \mid uv \in E(G)\}$ denote the \emph{arcs} of a graph~$G$.

\begin{construction}
    \label{con:psi-to-wsat}
    Fix an instance $(G,H, \mathcal{P})$ of \psgi{}. We construct a formula $\varphi$, with $\mathcal{V}(\varphi) \coloneqq V(G) \cup E(G)$, as follows:
    \begin{align*}
        \varphi \coloneqq \bigwedge_{x \in V(H)} \left( \bigvee_{v \in P_x} v \right) \quad \land\quad \bigwedge_{\overrightarrow{xy} \in A(H),\,v \in P_x} \left( v \Rightarrow \bigvee_{u \in N_G(v) \cap P_y} uv \right).
    \end{align*}
\end{construction}

Intuitively, setting a variable in $\mathcal{V}(\varphi)$ to true indicates that the corresponding vertex or edge is selected as part of the subgraph isomorphism. The first part of the conjunction ensures that for every vertex $x \in V(H)$, at least one vertex $v$ is selected from the set $P_x$. The second part of the conjunction guarantees that for every arc $\overrightarrow{xy} \in A(H)$, the subgraph includes a corresponding edge from a selected vertex in $P_x$ to some vertex in $P_y$. By setting $k \coloneqq |V(H)| + |E(H)|$ to accommodate for the additional variables required to satisfy the formula, we ensure that exactly the vertices and edges corresponding to a valid subgraph isomorphism are selected, resulting in a correct reduction from \psgi{} to \wsat{}.

For \cref{sec:stcon}, we need \wsat{} with $(1,1)$-bounded variables, whereas variable occurrences are unbounded in \cref{con:psi-to-wsat}. To address this, we modify the formula by splitting up the individual variables:

\begin{construction}
    \label{con:psi-to-wsat3}
    Fix an instance $(G,H, \mathcal{P})$ of \psgi{}. We construct a formula $\varphi$, with $\mathcal{V}(\varphi) \coloneqq \{ v_y\ |\ \overrightarrow{xy} \in A(H),\,v \in P_x \} \cup E(G) \cup A(G)$, as follows:
    \begin{align*}
        \varphi \coloneqq& \bigwedge_{x \in V(H)} \left( \bigvee_{v \in P_x} \left( \bigwedge_{y \in N_H(x)} v_y \right) \right) 
         \land \bigwedge_{\overrightarrow{xy} \in A(H),\,v \in P_x} \left( v_y \Rightarrow \bigvee_{u \in N_G(v) \cap P_y} \overrightarrow{vu} \right)\\
        & \land \bigwedge_{uv \in E(G)} \left((\overrightarrow{vu} \lor \overrightarrow{uv}) \Rightarrow uv\right)
    \end{align*}
\end{construction}

For each $x \in V(H)$, choosing a vertex $v \in P_x$ corresponds to setting all variables $v_y$ to true for each neighbor $y \in N_H(x)$.
An edge $uv \in E(G)$ is selected if at least one of the arc variables $\overrightarrow{uv}$ or $\overrightarrow{vu}$ is set to true, which in turn forces the edge variable $uv$ to be true.
Intuitively, by setting $k \coloneqq 5|E(H)|$, we ensure that exactly the vertices and edges corresponding to a valid subgraph isomorphism are selected.
After transforming $\varphi$ into NNF, we obtain a \emph{$3$-normalized formula}, that is, a conjunction of disjunctions of conjunctions of literals. In NNF, the variables become $(1,1)$-bounded\footnote{Recall that each implication $A \Rightarrow B$ is replaced by $\neg A \lor B$ in NNF, which negates the left-hand side.}---each variable $v_y$ occurs once positively in the first part of the conjunction and once negatively in the second; each arc variable $\overrightarrow{vu}$ occurs once positively in the second and once negatively in the third; and each edge variable $uv$ occurs exactly once, positively, in the third part.
This gives the following theorem, which answers the question of Kanj and Szeider mentioned above:

\begin{theorem}
    \label{thm:psi-to-wsat-11}
    \wsat{} restricted to $3$-normalized formulas and $(1,1)$-bounded variables is \wone{}-hard parameterized by $k$. Moreover, there is a parameter-linear reduction from \psgi{} to it.
\end{theorem}
\begin{proof}
We first aim to show that there exists a subgraph of $G$ isomorphic to $H$ while respecting the partition $\mathcal{P}$ if and only if there exists a truth assignment $I$ such that $I(\varphi) = 1$ and $w(I) \leq k = 5 |E(H)|$. We will prove both directions separately.

\begin{description}
    \item[$(\Rightarrow)$] Suppose there exists an injective mapping $\phi\colon V(H) \to V(G)$ such that for every edge $xy \in E(H)$, $\phi(x)\phi(y) \in E(G)$, and for every vertex $x \in V(H)$, we have $\phi(x) \in P_x$. We construct a truth assignment $I$ where for every vertex $x \in V(H)$ and neighbor $y \in N_H(x)$, we have $I(\phi(x)_y) = 1$; for every edge $xy \in E(H)$ with $u \coloneqq \phi(x)$ and $v \coloneqq \phi(y)$, we set $I(uv) = I(\overrightarrow{uv}) = I(\overrightarrow{vu}) = 1$. All other variables are set to false. By this construction, we have $w(I) = \sum_{x \in V(H)} \deg_H(x) + 3|E(H)| = 5|E(H)| \leq k$. It remains to be shown that $I(\varphi) = 1$.

    For the first part of the conjunction, consider an arbitrary vertex $x \in V(H)$. Then for vertex $v = \phi(x) \in P_x$ and all $y \in N_H(x)$, we have $I(v_y) = 1$. Thus, each conjunct of the first part is satisfied.

    For the second part of the conjunction, consider an arbitrary vertex $x \in V(H)$, one of its neighbors $y$ in $H$, and an arbitrary vertex $v \in P_x$. If $v \not= \phi(x)$, then $I(v_y) = 0$ and the implication is vacuously true. If $v = \phi(x)$, observe that for edge $xy \in E(H)$, we must have $\phi(x)\phi(y) \in E(G)$. Thus, for $u = \phi(y) \in N_G(v) \cap P_y$, we must have $I(\overrightarrow{vu}) = 1$ by construction, which satisfies the disjunction on the right side of the implication. Thus, all implications in the second part are satisfied.

    Since the implications in the third part are all trivially satisfied by construction, we must have that all conjuncts of $\varphi$ are satisfied, and we conclude that $I(\varphi) = 1$.

    \item[$(\Leftarrow)$] Suppose there exists a truth assignment $I$ such that $I(\varphi) = 1$ and $w(I) \leq k$. We begin by analyzing the structure of $I$, counting the variables set to true in each group separately:
    \begin{itemize}
        \item \emph{Vertex variables:} For all conjuncts of the first part to be satisfied, we must have that for each $x \in V(H)$, at least one vertex $v \in P_x$ has all its corresponding vertex variables $v_y$ set to true, amounting to at least $\sum_{x \in V(H)} \deg_H(x) = 2|E(H)|$ variables set to true.
        \item \emph{Arc variables:} Each of these vertex variables is the left side of an implication in the second part of the conjunction, requiring the right side to be satisfied as well. Since the right sides of distinct implications share no arc variables, this requires at least $2|E(H)|$ arc variables set to true.
        \item \emph{Edge variables:} Each arc variable set to true forces the corresponding edge variable to be true to satisfy the implications in the third part. Since each edge variable $uv$ is forced only by $\overrightarrow{uv}$ and $\overrightarrow{vu}$, this requires at least $|E(H)|$ edge variables set to true.
    \end{itemize}
    This amounts to a minimum of $5|E(H)|$ variables set to true in $I$. However, we know that $w(I) \leq k = 5|E(H)|$. Thus, this lower bound is tight, and so is each of the three bounds above. Consequently, for each $x \in V(H)$, exactly one vertex $v \in P_x$ can have all its corresponding vertex variables $v_y$ set to true, and no other vertex variables are set to true. We now define a mapping $\phi\colon V(H) \to V(G)$ such that for each $x \in V(H)$, we have $\phi(x)$ mapped to this vertex $v$.

    Clearly, for every vertex $x \in V(H)$, we have $\phi(x) \in P_x$ by construction. It remains to be shown that for every edge $xy \in E(H)$, the edge $\phi(x)\phi(y)$ exists in $E(G)$. To this end, consider an arbitrary edge $xy \in E(H)$ and let $v \coloneqq \phi(x)$. Since $v_y$ is set to true, some arc variable $\overrightarrow{vu}$ with $u \in P_y$ is set to true, and thus also the edge variable $uv$ by the third part. Due to the tight bound, $uv$ is set to true only if both $\overrightarrow{uv}$ and $\overrightarrow{vu}$ are, and each arc variable set to true is required by some vertex variable set to true. Since $v \in P_x$, the arc variable $\overrightarrow{uv}$ can only be required by $u_x$, so $u_x$ is set to true, which implies $u = \phi(y)$. Hence, $\phi(x)\phi(y) = vu \in E(G)$, completing the proof that $\phi$ constitutes a subgraph isomorphism.
\end{description}

Finally, $\varphi$ has $3|E(G)| + \sum_{x \in V(H)} |P_x| \cdot \deg_H(x) \le 3|E(G)| + |V(G)| \cdot |V(H)|$ variables, each occurring at most twice, so $\varphi$ can be constructed in polynomial time. Since $k = 5|E(H)|$, this is a parameter-linear reduction from \psgi{}. As \psgi{} is \wone-hard, so is \wsat{} restricted to $3$-normalized formulas and $(1,1)$-bounded variables.
\end{proof}

For \cref{sec:gcon}, we need \wsat{} on CNF formulas rather than $3$-normalized formulas. We show that \wsat{} remains \wone-hard on CNF formulas with $(2,1)$-bounded variables, again via a parameter-linear reduction from \psgi{}. To this end, we modify \cref{con:psi-to-wsat} by splitting up the vertex variables differently:
\begin{construction}
    \label{con:psi-to-wsat4}
     Fix an instance $(G,H, \mathcal{P})$ of \psgi{}. For each $x \in V(H)$, let $\pi[x]\colon N_H(x) \to \{1,2,\dots,\deg_H(x)\}$ be a bijection assigning a unique index to each neighbor $y \in N_H(x)$. We construct a formula $\varphi$ with $\mathcal{V}(\varphi) \coloneqq \{ v_i\ |\ x \in V(H),\, v \in P_x,\, 0 \le i \le \deg_H(x) \} \cup E(G)$ defined as:
    \begin{align*}
        \varphi \coloneqq& \bigwedge_{x \in V(H)} \left( \bigvee_{v \in P_x} v_0 \right) 
         \land \bigwedge_{\overrightarrow{xy} \in A(H),\,v \in P_x} \left( v_{\pi[x](y)} \Rightarrow \bigvee_{u \in N_G(v) \cap P_y} uv \right)\\
        & \land \bigwedge_{x \in V(H),\, v \in P_x,\, i < \deg_H(x)} \left( v_{i} \Rightarrow v_{i+1} \right)
    \end{align*}
\end{construction}

For each $x \in V(H)$, if vertex $v \in P_x$ is selected to be part of the subgraph isomorphism, then all variables $v_i$ for $i \le \deg_H(x)$ are set to true due to the chain of implications. By setting $k \coloneqq |V(H)| + 3 \cdot |E(H)|$ to accommodate for the additional variables required to satisfy the formula, we again ensure that exactly the vertices and edges corresponding to a valid subgraph isomorphism are selected. Observe that the transformation into NNF yields a CNF formula with $(2,1)$-bounded variables---variable $v_0$ occurs once positively in the first part of the conjunction and once negatively in the third; each variable $v_i$ for $i > 0$ occurs once negatively in the second part and at most once in each polarity in the third; and each edge variable $uv$ occurs twice positively in the second part. This leads to the following theorem:

\begin{theorem}
    \label{thm:wsat-gcs}
    \wsat{} restricted to CNF formulas and $(2,1)$-bounded variables is \wone{}-hard parameterized by $k$. Moreover, there is a parameter-linear reduction from \psgi{} to it.
\end{theorem}
\begin{proof}
We first aim to show that there exists a subgraph of $G$ isomorphic to $H$ while respecting the partition $\mathcal{P}$ if and only if there exists a truth assignment $I$ such that $I(\varphi) = 1$ and $w(I) \leq k = |V(H)| + 3 \cdot |E(H)|$. We will prove both directions separately.

\begin{description}
    \item[$(\Rightarrow)$] Suppose there exists an injective mapping $\phi\colon V(H) \to V(G)$ such that for every edge $xy \in E(H)$, $\phi(x)\phi(y) \in E(G)$, and for every vertex $x \in V(H)$, we have $\phi(x) \in P_x$. We construct a truth assignment $I$ where for every vertex $x \in V(H)$ and $i \le \deg_H(x)$, we have $I(\phi(x)_i) = 1$; for every edge $xy \in E(H)$, we set $I(\phi(x)\phi(y)) = 1$. All other variables are set to false. By this construction, we have $w(I) = \sum_{x \in V(H)} (\deg_H(x) + 1) + |E(H)| = |V(H)| + 3|E(H)| \leq k$. It remains to be shown that $I(\varphi) = 1$.

    For the first part of the conjunction, consider an arbitrary vertex $x \in V(H)$. Then for vertex $v = \phi(x) \in P_x$, we have $I(v_0) = 1$. Thus, each conjunct of the first part is satisfied.

    For the second part of the conjunction, consider an arbitrary vertex $x \in V(H)$, one of its neighbors $y$ in $H$, and an arbitrary vertex $v \in P_x$. If $v \not= \phi(x)$, then $I(v_{\pi[x](y)}) = 0$ and the implication is vacuously true. If $v = \phi(x)$, observe that for edge $xy \in E(H)$, we must have $\phi(x)\phi(y) \in E(G)$. Thus, for $u = \phi(y) \in N_G(v) \cap P_y$, we must have $I(uv) = 1$ by construction, which satisfies the disjunction on the right side of the implication. Thus, all implications in the second part are satisfied.

    Since the implications in the third part are all trivially satisfied by construction, we must have that all conjuncts of $\varphi$ are satisfied, and we conclude that $I(\varphi) = 1$.

    \item[$(\Leftarrow)$] Suppose there exists a truth assignment $I$ such that $I(\varphi) = 1$ and $w(I) \leq k$. We begin by analyzing the structure of $I$, counting the variables set to true in each group separately:
    \begin{itemize}
        \item \emph{Vertex variables:} For all conjuncts of the first part to be satisfied, we must have that for each $x \in V(H)$, the vertex variable $v_0$ for at least one vertex $v \in P_x$ is set to true. For each such variable, the chain of implications in the third part requires that all variables $v_1, \dots, v_{\deg_H(x)}$ are set to true as well, amounting to at least $\sum_{x \in V(H)} (\deg_H(x) + 1) = |V(H)| + 2|E(H)|$ variables set to true.
        \item \emph{Edge variables:} Each of the $2|E(H)|$ variables $v_1, \dots, v_{\deg_H(x)}$ above is the left side of an implication in the second part of the conjunction, requiring the right side to be satisfied as well. Since each edge variable appears in the right sides of at most two implications, this requires at least $|E(H)|$ edge variables set to true.
    \end{itemize}
    This amounts to a minimum of $|V(H)| + 3|E(H)|$ variables set to true in $I$. However, we know that $w(I) \leq k = |V(H)| + 3|E(H)|$. Thus, this lower bound is tight, and so are both bounds above. Consequently, for each $x \in V(H)$, exactly one vertex $v \in P_x$ can have all its corresponding vertex variables set to true, and no other vertex variables are set to true. We now define a mapping $\phi\colon V(H) \to V(G)$ such that for each $x \in V(H)$, we have $\phi(x)$ mapped to this vertex $v$.

    Clearly, for every vertex $x \in V(H)$, we have $\phi(x) \in P_x$ by construction. It remains to be shown that for every edge $xy \in E(H)$, the edge $\phi(x)\phi(y)$ exists in $E(G)$. To this end, consider an arbitrary edge $xy \in E(H)$ and let $v \coloneqq \phi(x)$. Since $v_{\pi[x](y)}$ is set to true, some edge variable $uv$ with $u \in P_y$ must be set to true to satisfy the implications in the second part of the conjunction. Due to the tight bound, each edge variable set to true satisfies the right sides of two implications whose left sides are true. For $uv$, these are the implications of $v_{\pi[x](y)}$ and $u_{\pi[y](x)}$, so $u_{\pi[y](x)}$ is set to true, which implies $u = \phi(y)$. Hence, $uv = \phi(x)\phi(y) \in E(G)$, completing the proof that $\phi$ constitutes a subgraph isomorphism.
\end{description}

Finally, $\varphi$ has $|E(G)| + \sum_{x \in V(H)} |P_x| \cdot (\deg_H(x) + 1) \le |E(G)| + |V(G)| \cdot |V(H)|$ variables, each occurring at most three times, so $\varphi$ can be constructed in polynomial time. Moreover, we may assume that $H$ has no isolated vertices, as removing them does not affect the existence of a solution. Hence, $|V(H)| \le 2|E(H)|$, and thus $k = |V(H)| + 3|E(H)| \le 5|E(H)|$, so this is a parameter-linear reduction from \psgi{}. As \psgi{} is \wone-hard, so is \wsat{} restricted to CNF formulas and $(2,1)$-bounded variables.
\end{proof} %
\section{Conclusion}
\label{sec:conclusion}

We studied the complexity of configuring switch graphs to achieve connectivity\textemdash{}both local and global\textemdash{}under various structural restrictions, in both weighted and unweighted settings.
While we showed that the problem is intractable for many restricted classes of switches, we also identified structural conditions that permit efficient algorithms.

For the \textsc{On-Off} variants, the trivial algorithm that tries all sets of at most $\lambda$ switches to turn on runs in $|I|^{\lambda + \bigO(1)}$ time, and our \wone-hardness results show that this is unlikely to be improved to an \fpt{} algorithm, even for very restricted switches.
How close to optimal the trivial algorithm is depends on the complexity of \psgi{}: since our reductions are parameter-linear, any running-time lower bound for \psgi{} carries over.
Under \ethy{}, this would rule out $f(\lambda) \cdot |I|^{o(\lambda / \log \lambda)}$-time algorithms, nearly matching the trivial algorithm.
Under the recently proposed $k \times k$ Clique Hypothesis~\cite{Szeider26}, a bound of $f(\lambda) \cdot |I|^{o(\sqrt{\lambda})}$ still holds.
Which lower bound for \psgi{} holds under well-established assumptions, and thus whether the trivial algorithm is essentially optimal, remains an interesting open question.

Further, it is unknown whether \oolcs{} restricted to $\{1,1\}$-bounded connected switches is solvable in polynomial time, or perhaps in \fpt{}-time with respect to the solution weight, and, more generally, whether \wlcs\ is solvable in polynomial time for $\{1^\star\}$-bounded complete multipartite switches (as is the case for \lcs).
Beyond this, it is interesting to systematically investigate other properties, such as disconnecting two given terminals, and restrictions on the structure of the graph obtained by taking the union of all switch positions, such as tree-likeness; preliminary results in this direction are available~\cite{Kern25}.

Another direction is the influence of planarity on the complexity of connectivity switching problems.
Our reductions from \wsat{} show (with minor modifications, such as splitting high-degree vertices) that \oolcs{} and \oogcs{} remain \wone-hard parameterized by~$\lambda$ in planar switch graphs, even for constant switch size and maximum vertex degree.
However, we do not know how to additionally obtain connected switches without destroying planarity.
Is \oolcs{} on planar switch graphs with connected switches fixed-parameter tractable when parameterized by~$\lambda$ for constant switch size and maximum vertex degree?

\ifjournal
\section*{Declaration of generative AI and AI-assisted technologies in the manuscript preparation process}
\else
\section*{Use of Generative AI}
\fi

All research ideas, results, and proofs in this work were developed by the authors without the use of generative AI. During the preparation of the manuscript, the authors used GitHub Copilot and Claude Code (Anthropic) to assist with LaTeX typesetting and formatting, restructuring the presentation of existing arguments, identifying and correcting minor errors and gaps (such as typos, inconsistent notation, and routine proof details), and improving the wording of the text. After using these tools, the authors reviewed and edited the content as needed and take full responsibility for the content of the published article.

\bibliographystyle{plainurl}

\end{document}